**2018.04.28**

# Astrometry history: Hipparcos from 1964 to 1980

## Erik Høg

Niels Bohr Institute, Copenhagen University, Denmark     _ehoeg@hotmail.dk_

**Abstract:** Here follow three reports covering different aspects of the early history from 1964 to 1980 of the Hipparcos satellite mission. The first report "Interviews about the creation of Hipparcos" contains interviews from 2017 with scientists about how the mission was conceived up to the begin of technical development. The second report "From TYCHO to Hipparcos 1975 to 1979" is about the Hipparcos development based on  new material from my archive. From my 65 years dedicated to the development of astrometry, I argue that very special historical circumstances in Europe were decisive for the idea of space astrometry to become reality: Hipparcos did not just come because astrophysicists needed the data. The third report "Miraculous 1980 for Hipparcos" documents how the approval of the astrometric mission in January 1980 in competition with an astrophysical mission was only achieved with very great difficulty, even after an outstanding astrophysicist had presented overwhelming arguments that the astrometric mission would be scientifically much more important.

## CONTENTS

With these three reports I have done as promised in 2011 in http://www.astro.ku.dk/~erik/History.pdf : "_Further instalments in preparation:  On the Hipparcos mission studies 1975-79 and on the Hipparcos archives._"  Now in 2018, one month before my 86th birthday, **I do not have any further instalments in preparation**.

## _With best regards        Erik_                    http://www.astro.ku.dk/~erik

http://www.nbi.ku.dk/english/namely_names/2017/at-the-age-of-85-year-old-erik-hoeg-is-planning-a-satellite-to-be-launched-in-20-years/

# Overview with links to the individual reports

No. 1 -  2018.04.10:
# Interviews about the creation of Hipparcos

**Abstract:** Hipparcos was the first satellite to obtain "high-precision global absolute astrometry from space", and Gaia is the second spacecraft to do so. These satellites provide very strong astrometric foundation for all branches of astrophysics from the solar system to quasars, a foundation which must be kept up to date as astrophysics is rapidly developing, and it can only be done by such satellites. It is therefore interesting, for me even scaring, to think of how much the creation of Hipparcos leading to the approval in 1980 depended on a handful of astronomers during the preceding fifty years. This dependence is a fact for me who has witnessed and taken active part in the development of astrometry since 1953. This report contains interviews with a dozen persons on this question. Herewith I try to convey the evidence for this dependence as I have done in numerous reports before. By now, "astrometry has been regained" from a state of weakening or slow improvement before Hipparcos, and with Gaia, astrometry has even become an "almost respectable pursuit" as a colleague has said. This must be maintained in the future for the sake of astronomy and astrophysics. New information is given in an appendix about work in the Strasbourg Observatory on photoelectric astrometry in the 1950s and on satellite astrometry later on.

http://www.astro.ku.dk/~erik/xx/HipCreation.pdf

No. 2 - 2018.04.15:
# From TYCHO to Hipparcos 1975 to 1979

**Abstract:** With this report I follow the encouragement from several colleagues to my previous historical reports, they urged me to write more about these old times. I am going as deep as possible by means of documents, my own memory and in discussions by email with colleagues. - It is worth investigating how the Hipparcos astrometric satellite mission came about: was it almost a historical necessity that had to happen because astrophysicists needed the accurate positions, parallaxes and proper motion for the study of the Galaxy and the Universe and because the technical tools were available and affordable? Was it such general circumstances or did other more special historical circumstances play a major and even decisive role? My experience from 65 years dedicated to the development of astrometry shows me that special historical circumstances were needed and decisive.

http://www.astro.ku.dk/~erik/xx/Hip1975.pdf

No. 3 - 2017.12.15:
# Miraculous 1980 for Hipparcos
**Abstract:** Many astrophysicists would agree that the astrometric foundation of astrophysics with positions, motions and distances of stars is important in all parts of astronomy and astrophysics. But in a situation where they have to chose between an astrometric and an astrophysical project the majority will chose the astrophysical, even after an outstanding astrophysicist has presented overwhelming arguments that the astrometric mission would be scientifically much more important. This extremely challenging situation became real at meetings in ESA on 23/24 January 1980 when Hipparcos stood against an EXUV project. This appears in detail from new documents which also show how a majority for Hipparcos was nevertheless gathered. Other documents from before 1980 on space astrometry are briefly described by Jean Kovalevsky, Lennart Lindegren and the present author (called EH hereafter) and links are given.

http://www.astro.ku.dk/~erik/xx/HipApproval5.pdf

**2018.04.10**

*#1:The first of three reports on the early history of Hipparcos from 1964 to 1980*

## Interviews about the creation of Hipparcos

Erik Høg, Niels Bohr Institute, Copenhagen University, Denmark     *ehoeg@hotmail.dk*

**Abstract:** Hipparcos was the first satellite to obtain "high-precision global absolute astrometry from space", and Gaia is the second spacecraft to do so. These satellites provide very strong astrometric foundation for all branches of astrophysics from the solar system to quasars, a foundation which must be kept up to date as astrophysics is rapidly developing, and it can only be done by such satellites. It is therefore interesting, for me even scaring, to think of how much the creation of Hipparcos leading to the approval in 1980 depended on a handful of astronomers during the preceding fifty years. This dependence is a fact for me who has witnessed and taken active part in the development of astrometry since 1953. This report contains interviews with a dozen persons on this question. Herewith I try to convey the evidence for this dependence as I have done in numerous reports before. By now, "astrometry has been regained" from a state of weakening or slow improvement before Hipparcos, and with Gaia, astrometry has even become an "almost respectable pursuit" as a colleague has said. This must be maintained in the future for the sake of astronomy and astrophysics. New information is given in an appendix about work in the Strasbourg Observatory on photoelectric astrometry in the 1950s and on satellite astrometry later on.

## 1  Introduction

This report was born out of a statement by Michael Perryman on 30 August 2017 at the meeting in Lund (Lund 2017) to mark the retirement of Lennart Lindegren. In his brilliant and very fitting "Overview of Lennart´s contribution to science", Michael said: "No one can really say whether Hipparcos and Gaia would have existed without Lennart… …but we can say that they would have been very different missions without him". I immediately objected to the words "no one can really say..." because I had explained years ago how indispensable Lennart has been (Høg 2008b, 2011b). References to my other papers especially on the history of astrometry are contained in Høg (2018b), a list with about 50 items.

Further discussions at the meeting in Lund led me to send the following mail, at first to a historian I know quite well and soon after to 15 colleagues with relations to Hipparcos or Gaia. The interviews are collected in this report and they show how differently we look at such aspects of scientific endeavor. I believe we should have these answers in mind for the future when we argue for a Gaia successor mission to fly in twenty years. Some answers show great optimism about the future of astrometry, others much more scepticism.  We are here only speaking of "high-precision global absolute astrometry from space", i.e. the kind of astrometry obtained by Hipparcos and Gaia.

Ground-based astrometry without Hipparcos would probably have developed almost as it did. Without me, photon-counting astrometry would have been invented some years later than by Høg (1960), see the overview in Sect. 3 of Høg (2014), automatic meridian circles would have come later than 1984 and perhaps have by-passed the photon counting method and gone directly to use CCD scanning. Astrometry with large photographic plates and with CCDs would have come, though not with the good absolute reference system in the Hipparcos and Tycho-2 catalogues.

The mail as it was sent:

I have a question to you about the history of astrometry, a question I am asking a few colleagues. Please read the following and let me then know your opinion.

In my paper:
Høg E. 2011b, **Astrometry lost and regained.**
    Baltic Astronomy, Vol. 20, 221-230, 2011.
    http://esoads.eso.org/abs/2011BaltA..20..221H

I claim in Sec. 5 that the progress leading to the Hipparcos satellite was critically dependent on seven persons. **If anyone of them had been missing nobody could have filled his place in this particular development from 1925 to 1980.**

Most briefly the claim is made in the caption of Fig. 5, but please look critically at my discussion in Sec. 5. Am I right as seen by you???

Fig. 5. The development of photoelectric astrometry since 1925 and of the Hipparcos project was critically dependent on every one of the first six of these astronomers up to the approval in 1980. The seventh, Edward van den Heuvel, strongly advocated Hipparcos in the ESA decision process in 1980 although he himself as an X-ray astronomer had a direct interest in the competing EXUV mission.

## 2 An analysis of the interviews

In several responses I am being encouraged to write about these matters and I will do so. But only few (Kovalevsky, Egret, van Altena, van den Heuvel, Bastian, Mignard) seem to agree with my claim in Fig. 5, it is however not clear whether they agree that there was a kind of chain. Perhaps a formulation like, "given the state of the field and recruitment and funding mechanisms..." should have been included, but these conditions appear from my discussion in the paper. If the formulation had included "it is very unlikely that anybody could have filled his place..." it would probably have been more acceptable, but I am writing "I am sure" from my personal knowledge of the scene in those years before 1980.

The list of seven persons is not meant to be a complete list of key persons as some responses suggest, and one, Claus Fabricius on 24 Sep., thinks that this list almost implies "that everybody else made only insignificant contributions that anyone could have made". This is far from my intention which is to show that these persons were links in a chain over 50 years. Claus also thinks that I give "a very distorted view of the history" by including Bengt Strömgren and Otto Heckmann since they belong only in my personal scientific development (cf. Høg 2014, 2017a, b, c). But without Strömgren I would not have been on the scene at all, I could not have published Høg (1960) with the photon counting astrometry and with the inclined slits called "une grille de Høg" in France in those years and which were presumably important for Pierre Lacroute's great vision of space astrometry.

Photoelectric astrometry was intensely studied in Strasbourg in the mid-1950s by Pierre Bacchus the student of Pierre Lacroute director of the observatory. The thesis of Bacchus (Bacchus 1959) of 66 pages, I have only found now, on 16 October 2017, led by the report Kovalevsky (2009) where it is mentioned but without a reference. On this basis Lacroute could have seen that the new digital astrometric method in Høg (1960) was suitable for use in a scanning satellite. Jean Kovalevsky agrees with me that Lacroute might have started his work on space astrometry at this time, but we have no direct evidence on when it happened except that his first presentation of the ideas was in June 1965 at a colloquium in Bordeaux Observatory. The appendix discusses this question and gives more information about the work in Strasbourg about 1960.

**The new report Høg (2018) documents my design in 1975/76 of a scanning astrometry satellite with ten new features later implemented in the final Hipparcos mission. It appears from the historical context that this self-consistent design could not have been proposed at that crucial time by anyone else than me, in accordance with my claim above.**

My immediate answer to Claus Fabricius is given on 24 Sep. I agree with him that astrometry has not been "regained" forever, but that astrometry has become an "almost respectable pursuit". Astrometry has by the Hipparcos mission been regained from a state of weakening or slow improvement before Hipparcos, and even more by Gaia. This must be maintained in the future for the sake of astronomy and astrophysics.

Several responses do not want to engage in speculations of "what would have happened if something was missing...?", but Lennart rightly calls it **"a way to highlight the importance of certain things."**

Finally, although it really belongs first, Pierre Lacroute as the father of space astrometry and Jean Kovalevsky as his strong supporter were irreplaceable. I greatly admire Lacroute for his insistence in this matter during all the years from about 1964 up to 1975 when ESA was engaged, see Kovalevsky (2009). Lacroute was not an instrument designer at all, but he had a great vision and he tried hard.

## 3  The 12 interviews

A collection of slightly edited mails with answers is listed here in strict chronological sequence of each person's first reply.

### John L. Heilbron

I have at first asked John L. Heilbron, professor of history and the history of science emeritus, University of California in Berkeley about his opinion. He answered as follows, in fact within hours of my mail on 5 Sep. and our correspondence continued for a couple of days, to my great pleasure and education.

**Dear Erik,**   on 5 Sep.

The question is really not an historical one, as it is of the form, "what would have happened if something was missing...?" No doubt, the field would not have developed as it did without the players you mention as indispensable. But that is a truism: change any detail and the outcome will be different, although the difference might not be consequential.

The story you tell makes it plausible that without all of you the field might have declined or perished. Would / could someone else have arisen to take the place of one or another of you without sensibly affecting outcomes?  We cannot know. Perhaps a formulation like, "given the state of the field and recruitment and funding mechanisms, it is very unlikely that..." would be best.

I look forward to seeing you next month.

With warm greetings,

John

On 6 Sep 2017, at 07:04, Erik Høg wrote:

**Hi John**,

Thank you very much for your answer which will be a great help in my further discussion. In a coming note with such a discussion I would like to quote you in verbatim, if you agree?

Your last phrase is very good, just the way I am thinking. You wrote: ...a formulation like, "given the state of the field and recruitment and funding mechanisms, it is very unlikely that…" would be best.

My intention is to argue further on my case in which I strongly believe. I hope thus to draw the attention of historians of science to this case and urge them to find similar ones where "A few scientists were part of a chain of actions and events which brought a great progress of a branch of science. If any one of these scientists had been missing this progress would most probably not have happened, given the state of the field and recruitment and funding mechanisms."

Normally in science, big progress is obtained by scientists in teams. They work to stay in the frontline, or even to keep the leadership in Europe, or perhaps in the world. They know however that the progress would be obtained, if they were absent, just somewhat later. There are so many bright colleagues in the field that one of them could be missing without other loss than just some length of time, which is of course also important. Martin Rees, Astronomer Royal, has written in this sense in one of his books, but I cannot find the reference at this moment. He said that the great development of physics in the early 1900s would have happened even without Einstein, except perhaps for General Relativity which is truely Einstein's. But even that is not obvious I think, when you consider how close Hilbert was to GR at the same time as Einstein in November 1915, see p.141 in C. Reid (1970): Hilbert.

The case I argue about is very different. Astrometry was a weakening but important branch of astronomy from about 1900 when astrophysics was rapidly advancing. This is what I explain in Høg (2011b), "Astrometry lost and regained". (Few of my colleagues ever saw the allusion to John Milton's great poems of 1667 and 1671 which I read a few years ago.) The development of space astrometry took place in Europe, because only there were the good ideas and the human resources to support a big astrometry project as Hipparcos, a project costing about 600 million Euro to ESA plus several 100 millions to ESA countries for the data reduction. The competition is very hard to get such funding and if you don't get it you get nothing. I show that seven named person acted in a kind of chain: if one of them had been missing Hipparcos would most probably not have been approved in 1980, and probably never.

**Hi Erik**,

I doubt that my name or affiliation would have any positive effect, especially since I know nothing about the topic except what I have learned from you. You are welcome to quote any phrases from my email you find useful but whatever you say will be the stronger for not mentioning so doubtful an authority as I would seem (and am!).

**Hi Eric**,   on 7 Sep

Here is what I think: Einstein invented the SR and the GR; without him very probably some functional equivalent of the theories would have been developed; but not Einstein's theories. The special mind set of an Einstein or a Bohr gives a colour to their work and its interpretation that would not be duplicated merely by arriving at their equations by another route. "Results" are not univocal. Just think of the merry confusion that Bohr sowed with his peculiar way of regarding his atomic theory. Or the "simultaneous" discoveries of the conservation of energy, which, on close examination, turns out to have been a set of closely related but not identical theories.

If I understand you correctly, the aspect that gives the chain its uniqueness for you is that it managed to rescue a declining research specialty in a very big and unexpected way. I think that you have made your case that the rescue was a remarkable performance, against the odds, unforeseen, astonishing in the prevailing climate, requiring intimate teamwork, special skills, costly equipment, etc.; why try to add a proof of the unprovable, viz., that nothing like the project would have come into existence if any of your seven actors were replaced by someone else?

The only thing I am sure about is that my opinion in the matter has no probative value.

It appears that we may have something to talk about in Copenhagen!

Med venlig hilsen,

John

**Hi John,**

But your opinion is very important as you are an historian of science.

Especially interesting would be to hear if my further arguments are reasonable? and have they had an effect on your opinion?

**Hi Erik,**  on 10 Sep.

Of course I applaud not only your statement,

> I do think it is of some historical interest to pass on my intimate knowledge of the scene of astrometry in the interval from 1925 to 1980 in order to show how much the development depended on very few individuals. Clearly this is my personal view of the events but perhaps of some interest for my colleagues and for historians.

but also the seriousness with which you have gone about the task.

**Jean Kovalevsky** on 7 Sept.

Dear Eric

... ... Thank you  for your long explanation about the seven people you mention that were major contributors to Hipparcos. I fully agree with your choice.

**Jean Kovalevsky** again on 11 Sept.

Do not get embarrassed, it is a pleasure to have such discussions, so continue as much as you like. In my present position, living in an old people residence, there is nobody with whom to discuss scientific matters.

I read carefully your section 5. May I say that I do not concur with your pessimism. Astrophysicists, even at that time, and much more later, lacked terribly precise parallaxes for many types of stars in order to understand their behavior. They understood what space could do for them and already showed it in Frascati. From then on, they would insist more and more on doing something in this direction.

You say rightly that astrometry on ground was getting weaker. But does one need to be an astrometrist to build a Hipparcos or something similar? I do not believe so: it was built by engineers of ESA, MATRA, etc.

• To promote it? Certainly not: when I first presented it in 1978 or so in an astrometric colloquium in Austria, almost everybody thought it was a mad idea; to the best, they feared for their future.

• To perform the computation? Only partly, one needed essentially good specialists in computing sciences (in Fast there were no more than 3 or 4 people who made some astrometric observations - 2 years as for me.)

• On the contrary, we needed - we had- - good specialists in Fundamental astronomy.

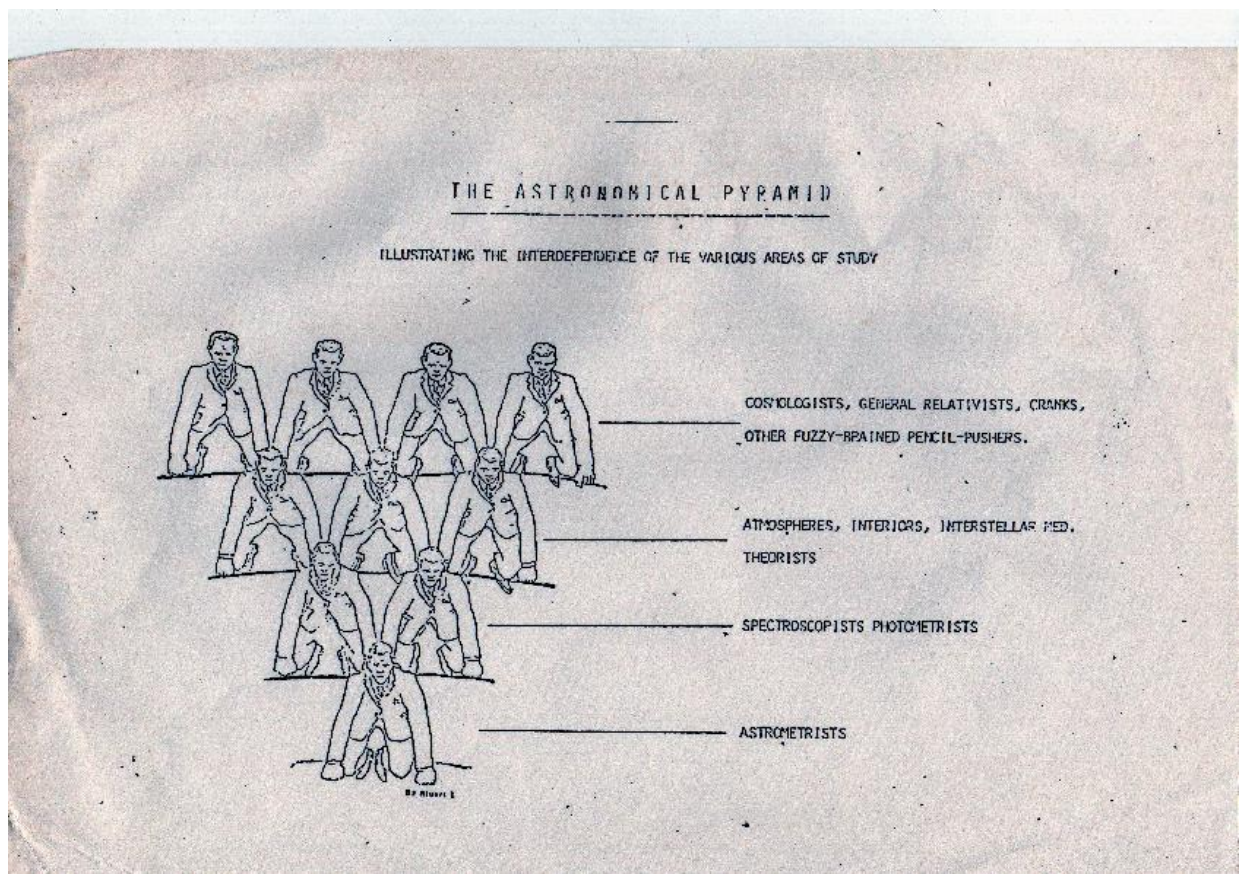

Figure 1  From Jean Kovalevsky, the astronomical pyramid: "a cartoon that I keep for many tens of years; people would not let the basis crash!" By Ron Probst 1974 for his astrometry master's thesis at McCormick Observatory, University of Virginia.

Would Hipparcos had not been chosen or failed, I am sure that there would have been a world-wide push for large production of astrometric data from space for the astrophysical community, as shown by the success of INCA. The success of Hipparcos showed even that they wanted much more. The adoption by ESA of GAIA is the result of a push by the astrophysical and cosmological communities and not astrometrical.

So, in my views and despite the wild imagination of Connes, something like Hipparcos had to be launched anyway, with or without astrometrists. I cannot imagine that Astronomy in general could survive without precise data on stars while all other techniques in all wavelengths were drastically improving. If not ESA, then NASA or the Japanese would have done it… Let me send you a cartoon that i keep for many tens of years; people would not let the basis crash!

Concerning Lacroute (Father of space astrometry, not Hipparcos, you are right) I was unaware of what XXX said you about the way Lacroute worked. I did not go to Strasbourg before 1975 or so, and I did not know of it.. [EH: This is answer to another mail to Jean from me before I knew the name of XXX, the name of the person is Bernard Traut, technical assistant at the observatory]

Finally, I admit that I have underestimated the role of van der Heuvel around 1980.

**Michael Perryman**  on 8 Sep.

I am traveling on some vacation and it will be a time before I can get to your mail. But more importantly, I do not want to get involved in personal perceptions of history. I have great respect for your views, and I think you should keep them as that - as your views. I cannot judge how history would have evolved in different circumstances and I do not want to speculate!

**EH** replied on 8 Sep. with an answer used in the introduction.

**Lennart Lindegren** on 10 Sept.

I very much appreciate your recording of the modern history of astrometry and the events that lead to the realisation of the Hipparcos mission. It must be immensely valuable for future historians to have a detailed account by somebody with first-hand knowledge of the events and persons involved, as well as a deep understanding of the technical issues. I also think the account of the actual facts is correct, as far as I know or remember them - although my memory is not as good as yours! Clearly these persons each had an important, even pivotal role in the process. However I feel uncomfortable about statements like what you write below: "If anyone of them had been missing...". Of course the developments would have been different and perhaps Hipparcos (and Gaia) would never had happened, or perhaps they had happened much later or in a very different form. But on the whole I do not find counterfactual speculations very useful, except as a way to highlight the importance of certain things.

Thanks for taking the time to write about these things!

**Jos de Bruijne** on 11 Sept.

Thanks for your message. It was nice meeting you in Lund two weeks ago! I appreciate and acknowledge that the 7 persons you mention have all played crucial roles in the history of astrometry. Unfortunately, predicting how the future would have developed in case one of these persons would not have been around at the time is so far from my area of expertise (space astrometry), that I would rather refrain from answering.

**Nigel Hambly** on 11 Sept.

Thanks for that. I can't really comment as my knowledge of the history of all this is lacking. Personally I'd be inclined to err on the side of understatement and point to key individuals as leaders rather than make bold statements about things not happening at all if any one had been absent…

**William van Altena** on 11 Sept.

I have just read your Sec. 5 and agree with your evaluation of the importance of the cited individuals in the launching and success of the Hipparcos project. Concerning the US, I would be even more pessimistic about the US/NASA launching an astrometric mission - SIM is an example. This is especially true for the current administration which is about as anti-science as is possible and is only interested in more riches for the already super rich and destroying any project that has a social conscience.

**Ed van den Heuvel** on 14 Sept.

I agree with what you write in the figure caption about the 7 persons.

**Ed van den Heuvel** on 7 Oct.

Dear Erik,

Thank you for your "Interviews about the creation of Hipparcos"

Reading all the interviews (emails) and thinking it over a bit further, I would like to expand a bit my reply of 14 september (which was very brief), as I think that I may not have been so irreplaceble as the others in the group of 7. After all, I played a role only on one day, 24 Januray 1980, and that was all.

Therefore, I would like to replace my statement of 14 September by the following one:

 "I largely may agree with what you write in the figure caption about the 7 persons, though with the following reservation: I think that if another astrophysicist had been asked by the AWG chair to compare, in the crucial 24 January 1980 AWG meeting, the Hipparcos and the EXUV missions, and if this astrophysicist would have looked carefully at which of these missions would be most important for astrophysics, he/she would have come to the same conclusion as I. So, I am not so sure that I have been as indispensable for the success of the Hipparcos mission as the other persons in the figure. Another astrophysicist in his/her right mind could very well have done the same job. But history made it such that I was asked to do this job, and I am very happy about how this came out."

**EH** comment on 15 Oct.

In an ideal world, another astrophysicist could have done as Edward van den Heuvel says, but Ed was not such an ideal astrophysicist, Ed was an X-ray astronomer and he had a big stake in the EXUV mission. He could not be expected to be impartial about the two mission. He was chosen by the AWG chairman expecting Ed to speak for the EXUV mission. The evidence in the reports EH2011 and EH2017d speaks for this assumption, and I know it is true from a witness who wants to be anonymous. Ed did not have a long lasting role in the creation of Hipparcos but his short role on one day in January 1980 was very critical even decisive. We are all happy how it went.

Many years later in 2006 when we happened to meet at the IAU Assembly in Prague, Ed told me the story now documented in the two reports. Ed presented the superior scientific capabilities of Hipparcos for the AWG and SAC, but did not trust that this alone would be enough. He also spoke with some members before the voting thus ensuring a majority of 8 to 5 for Hipparcos.

**Daniel Egret**  on 15 Sept.

**Hi Erik,**

That's a pleasure to receive a message from you, as this reminds me of the good ol' times of Hipparcos and Tycho.

I fully subscribe to your formulation.

The two other names I could have thought of, for France, are Jean Delhaye (who was critical for the French involvement in Hipparcos, but more on the political side,  and his background, as you know, was more stellar statistics than astrometry). And, in an earlier period (you mention : since 1925)  André Danjon  who  was essential  for  rebuilding  French  astronomy  after  the  war, and  driving  French observatories towards fundamental astronomy. But again, his role has been probably wider and more general, so his impact on the development of astrometry and Hipparcos is probably more indirect.

Best wishes for your book. I look forward !

All the best

Daniel   (currently astronomer emeritus in Observatoire de Paris)

**Hi Daniel**  on 15 Sept.

It is nice to hear from you and of course also that you agree with my idea of the key persons. Do you especially agree when I say that without Pierre Lacroute and Jean Kovalevsky there would have been no Hipparcos??? Or could anyone have replaced any of them??? We must think of real persons.

I consider Lacroute as the father of space astrometry, not of Hipparcos and Jean has just confirmed that he fully agrees with that view. I am having a very interesting correspondence with Jean and he urges me to continue since he is missing scientific discussions where he now lives, in a home for old people.

The two persons you mention were important but not as much as the ones in my list. I could mention Peter Naur my Danish mentor who meant so much for me, but I will not place him among the seven I mention. See more e.g. in: Høg (2017b).

My main objective is not to be complete, but to show a case of an important scientific development where people acted in a chain, if one had been missing the big goal would not have been reached. Do you know any other such chain in the history of science???

EH: The answer from Daniel Egret came later and is placed as Sect A4 in the appendix.

**Ulrich Bastian** on 18 Sept.

I can't really judge that list. That the six are indispensable is clear to me, but I have no view of which other ones might have been important at the same level. In other words, whether this list is "correct" in the sense of being complete at the given level of "importance".

**Timo Prusti** on 22 Sept

In the text you already quote that someone said you can never be sure what had happened if one person was missing from the chain. I tend to incline to that direction despite your reasoning in favour for this strong statement. In any case these developments took place before my time so I cannot really confirm or deny your claims. However, to me the key thing is the story why these 7 people to your opinion could not have been missed. Then based on that story you are free to make your claim and continue with the points you mention why the replaceability would not have worked. Thus I join the others encouraging you to continue writing down these historical events.

**Valeri Makarov** on 23 Sept.

Thank you for sharing this interesting paper that has been slipping below my radar, to be honest, as I have been fighting dragons for the past few years. I am not in a position to judge the relative contribution from the founders of modern astrometry, being myself a very late incomer, but I believe Lacroute definitely deserves his place among the seven. He indeed had a great, revolutionary vision. I also like the young spirit of enthusiasm and innovation that you managed to convey in the paper. These are rare commodities now, it seems. For you, apparently, the turning point was when the leader encouraged "free thinking". I suspect, good things do not come about without this.

**Claus Fabricius** on 24 Sept.

Regarding your question, I think it is pointless to enter into counterfactual speculations, so you are - in my opinion - asking the wrong question. What you, on the other hand, can do, is to say that the following persons - according to your judgement - played a particularly important role in the Hipparcos project, but without saying - as you almost do - that everybody else made only insignificant contributions that anyone could have made. Like Valeri, I was too little involved in the early phases of Hipparcos to be able to say who were the key persons, but I would definitely leave out, e.g., Strömgren and Heckmann. They may well have been important to your personal scientific development, but not to Hipparcos. Mixing these two groups gives a very distorted view of the history. The role of van den Heuvel is interesting in the context of the AWG decision, which in itself is worth exploring, but then you must discuss all members of the AWG, their views on astrometry, and which conflicting interests they had. Again, this is not Hipparcos itself.

As for the current situation for astrometry, it is characteristic that the core Gaia astrometry - with Sergei as an exception - is driven by retired people. Gaia has been a real game changer, so I am not so sure that astrometry has been "regained", but it has at least become an "almost respectable pursuit".

**Hi Claus,** on 24 Sept

I am very grateful for your text which very clearly draws my attention to misunderstandings in my original arguments. I will explain my point better in the planned report. But already here: the list of seven persons was never meant to be a complete list of key persons. It was meant to be a list of persons such that as I wrote:

> **If anyone of them had been missing nobody could have filled his place in this particular development from 1925 to 1980.**

The persons were links in a chain and this is my point. Such chains are probably rare in science and historians may not even believe in this chain concept at all. I will include a thorough correspondence with a historian  John L. Heilbron (professor emeritus, California University, Berkeley) who has this opinion, but nevertheless I am sure of my point.

In a new explanation I will perhaps begin with Lacroute and Kovalevsky. I will keep the seven named persons and only them but there could be others, in my own case there was Peter Naur without whom I would not have been able to digitize the meridian circle in Hamburg as I have explained in a recent report (Høg 2017b).

Strömgren is a link in the chain because of his experiments and papers in 1925, 1926 and 1933 which were in my mind all the time and because he ordered a new meridian circle in 1940 which was installed in Brorfelde in 1953. He picked me to work on that instrument in 1953 because he had no choice, pure chance it was and also that I had the talent and became fascinated by the instrument and by astrometry.

Heckmann is included because he immediately supported my idea in 1960, but his role could possibly have been taken also by his successor, Hans Haffner.

I agree with you that astrometry is at the moment not "regained forever". But it has at least become an "almost respectable pursuit".

This discussion is very useful for my understanding of how all these interviewed persons think and I will include it in verbatim in the new report. It will then be possible for readers to understand better this very important part of the history of astrometry, even if they do not accept my chain concept.

**Frederic Arenou** and **Catherine Turon** on 1 Oct.

Unfortunately, I am completely stuck inside the Gaia validation, unable to do anything else yet. Perhaps in a few months from now!

**Francois Mignard** on 1 Oct.

I agree fully with your selection of the key-people regarding the advent of Hipparcos. However to make it a great success I would add Michael Perryman, although he played no role in the selection, which is the main point of your paper.

From a discussion I had recently with J.C. Pecker it seems that the place of P. Bacchus has not been recognized at the right level, probably screened (unintentionally) by P. Lacroute.

**EH** answered: Correct, my paper is only about the time up to 1980. My point is only about the chain of the seven people mentioned, i.e. not about recognizing Bacchus. But a few years ago I looked for a few hours through the Lacroute archive in Meudon, a stack of paper only about 8 cm thick. One of the things I looked for specifically was the question of Bacchus's recognition and I found nothing indicating a screening, e.g. in the various obituaries was nothing indicating a greater role about the ideas on space astrometry, they are from Pierre Lacroute. I greatly admire Lacroute for his insistence in this matter during all these years from about 1964 up to 1975 when ESA engaged, see Kovalevsky (2009). He was not an instrument designer at all, but he had a great vision and he tried hard.

**Francois Mignard** on 7 Oct.

Correction of a possible misunderstanding: The word "screening" I used (not a good choice !) clearly meant unintentionally, and did not hint on a possible attempt to hide something on the side of Lacroute. I just hinted that Bacchus may have been "shadowed" by Lacroute, and this was a pure assumption, nothing to support the idea. - I hope this is clear now, and in case this was misunderstood I apologized for the confusion.

**Acknowledgements:** I am grateful to the respondents for their replies and for their permission to let these be included in this report, and to Sergey Klioner and Ron Probst for additional information on Figure 1.

# Appendix

## From Strasbourg around 1960 and about the early phases of Hipparcos

**Abstract:** The vision of astrometry from a scanning satellite was born in Strasbourg in the 1960s by Pierre Lacroute. His student Pierre Bacchus completed his thesis in 1959 about photoelectric measurement of double stars. This paper assumes of course analogue amplification of the photo current since this was the only available option in the 1950s. The paper by Høg (1960) introduced photon counting for astrometry as would become an option for astrometry with digital computers which were however still in their infancy. This method was adopted for space astrometry by Pierre Lacroute and I try below to find out when that happened and found so far that it happened between 1960 and 1965. In Sect. A2 follows an overview of the time 1975 to 1979 and as Sect. A3 a report from Jean-Louis Halbwachs of recent interviews with Mr. Bernard Traut, technical assistant at the observatory. A4 and A5 are letters by respectively Daniel Egret and Lennart Lindegren which came on 30 October, the day I had distributed the first draft of this appendix. Finally, a list of further references follows.

## A1   Photoelectric astrometry

By Erik Høg

At first, a quote from the above introduction, "Photoelectric astrometry was intensely studied in Strasbourg in the mid-1950s by Pierre Bacchus the student of Pierre Lacroute director of the observatory. The thesis, Thèse de Doctorat d'Etat, of Bacchus (Bacchus 1959) of 66 pages, I have only found now, on 16 October 2017, led by the report Kovalevsky (2009) where it is mentioned but without a reference. On this basis Lacroute could have seen that the new digital astrometric method in Høg (1960) was suitable for use in a scanning satellite. Jean Kovalevsky agrees with me that Lacroute might have started his work on space astrometry at this time, but we have no direct evidence on when it happened except that his first presentation of the ideas was in June 1965 at a colloquium in Bordeaux Observatory. The appendix discusses this question and gives more information about the work in Strasbourg about 1960."

The paper by Bacchus describes a modulation method and assumes of course analogue amplification of the photo current since this was the only available option in the 1950s. In space however a digital method of detection would be required, the photon counting astrometry.

According to Kovalevsky (2009) the first astronomers to present anything about astrometry in space were Couteau & Pecker (1964) at the Nice Observatory. They discuss double stars and planetary systems and conclude that the Moon would be the best platform. They recommend to use the photoelectric modulation techniques described by Bacchus i.e. with analogue amplification. This indicates that photon counting astrometry was not known or only little appreciated in Nice in 1964.

The important step from the analog techniques of Bacchus to the photon counting is mentioned only once by Kovalevsky on p.3 with the words "P. Lacroute [in the paper from 1966] considers that one should have at least 10 counts for each slit". Naturally so because in 1966 photon counting had become the only detection method to consider. Thus I have not seen any information on when Lacroute became aware of photon counting for astrometry, but I should believe it happened already in 1960 through my paper. With certainty we can only say it happened between 1960 and 1965.

My photon counting astrometry was publicly discussed soon after it was published. This happened twice in Astron. Journ. in 1961, first by van Herk & van Woerkom who doubted my method would work at all and later by myself in a reply. It was not presented at the IAU Assembly 1961 in Berkeley where I participated, but I was still too young and shy to speak up. I presented it at a meeting of the IAU Commission for astrometry in Hamburg in 1964 on invitation by the commission president Mr. Scott from USNO, and the instrument was shown at the meridian circle in the Bergedorf Sternwarte. I do not recall if anyone from France was present, but I believe Yves Requième from Bordeaux was there and presented his photoelectric meridian circle. He did not use photon counting but developed a mechanism centering on the star for which analog amplification was used. He told me once that the choice of this system was done on advice from Andre Danjon.

From those years I remember a reprint sent to me of a report probably by a student of Bacchus which spoke of "une grille de Høg" i.e. the V-shaped grid proposed in Høg (1960). This reprint is long lost from my files of paper, but it showed the interest in France for this feature. A "Høg's grid" was used by

Sauzeat (1974) and Crézé et al. (1982).

The annual meeting of French astronomers took place in Paris in June 1965 where I was invited to attend a "Colloque Astronomie Fondamentale et Mécanique Céleste" and gave two presentations: "photoelectric measurement of star transits" and "photoelectric reading of declination circle". This was my first invitation to another country than Denmark or Germany and all the other 17 presentations than mine in the two days meeting were in French as appears from the program recently received from my colleague through all the years, Yves Réquième from Bordeaux Observatory.

## A2  Overview of the Hipparcos project from 1975 to 1979

By Erik Høg

Pierre Lacroute presented his ideas of space astrometry in Prague in Lacroute (1967) where I heard him, but I had never spoken with him before we met in October 1975 in Paris for the first meeting of the ESA Mission Definition Group. This meeting changed me from being very sceptical about space astrometry to become enthusiastic as explained and documented in Sect. 4 of Høg (2011b).

For the next meeting in December 1975 I proposed a new design with one-dimensional scanning, active attitude stabilization, revolving scanning mode, input catalogue etc., in all seven new features are listed in Høg (2011b), but there were in fact ten new features as now listed in Sect. 3.1 of Høg (2018).

In January 1976, Pierre Lacroute invited me by the letter in Figure 1 of Høg (2018) to come to Observatoire de Paris before a study group meeting for a discussion between just us two, and he tells in the letter how much he appreciates that younger persons have now entered the project. We spent several hours in fruitful and pleasant exchange of ideas. Lacroute had immediately in December 1975 agreed to aim for a scanning satellite with one telescope, not with two telescopes as in the TD Option from Frascati, see ESRO (1975). The Spacelab option, the preferred option in the conclusions from Frascati, was still mentioned as a possibility in a note (see Høg 2018) in March 1976. He adopted an image-dissector tube (IDT) to become the very efficient primary detector behind a modulating grid instead of several photomultipliers behind long slits as in his original TD Option. In notes from December 1975 to March 1976 he also introduced a star mapper with slits and photomultipliers placed before the main field, able to detect stars and measure their position as required to point the IDT spot even without using an input catalogue. But his acceptance of other features in my proposal came more gradually and the chairmen of the group were obviously keen to avoid any decisions which might be premature or perhaps against Lacroute's ideas.

## A2.1  The development 1976 to 1979

I remember members of the study group saying that we were diverging because Lacroute maintained two-dimensional scanning with inclined slits, passive stabilization and no input catalogue since he considered this to be simpler and more safe. I assured them that we would soon converge towards my design.

But it took much longer than expected, nearly three years before the use of a modulation grid for two-dimensional scanning, a beam combiner in three parts, and passive scanning were definitively dropped. This appears from the fact that these options are still mentioned in ESA (1978), the "report on Phase A study", and in Barbieri & Bernacca (1979), the proceedings from the colloquium in Padua. Likewise according to the study report of 26 April 1978, the use of an input catalogue had not yet been decided. The final study report ESA (1979) does mention the input catalogue in Sect. 1.7.

The use of an input catalogue had, however, been decided already in a meeting of the Science Team on 23-24 November 1977 according to Høg (1997). I could therefore in December 1977 distribute an inquiry which I had held ready for some time on projects for the satellite mission to astronomers mainly in ESA countries and I went on a round trip giving lectures at a number of institutes to generate more interest and support for the mission project. This resulted in 80 projects of scientific investigation defined by about 50 astronomers at 12 institutions in ESA countries as reported in Høg (1979). These projects were analyzed by Høg (1979) and Turon Lacarrieu (1979) for their consequences on e.g. Galactic astrophysics.

The complete acceptance of one-dimensional scanning came after January 1979, according to Høg (1997 p. xxx). The same page contains further notes on the slow acceptance of the principles proposed in December 1975.

Lacroute's idea of attitude stabilization by the gravity gradient would conceivably work in the low-earth orbit originally assumed, but at the meeting of the science team in November 1977 we were directed by ESA to design the mission for a geostationary orbit which had become possible with the new ESA launcher Ariane. That killed the gravity gradient as an option for stabilization and active attitude control became obligatory. That also decided in favor of an input catalogue. According to Høg (1997), the main reason to keep the option with passive stabilization was its smooth rotation, while the active control with reaction wheels produced astrometric jitter of unknown size, and this question was never precisely answered. But the problem with reaction wheels was radically solved when MATRA introduced cold gas control. That came much later, during Phase B.

> The evolution of the Hipparcos project is described in ESA (1989). An account of the early phases of the Hipparcos project up to inclusion of the Tycho experiment in 1981 was presented in Venice as Høg (1997). The above text covers the period up to 1979 with a somewhat different emphasis. The early phase in the creation of Hipparcos and the mission selection process in 1980 are broadly described in Chapter 5 of the book by Perryman (2010). In Høg (2018) I am describing my archive from those years focusing on the instrument design.

## A3   Interviews with a technical assistant at Strasbourg Observatory

By Jean-Louis Halbwachs

From Jean-Louis Halbwachs at the Observatoire de Strasbourg I have received the following "overview of the activities in Strasbourg pertaining to the Hipparcos programme". It is based on his recent interviews with Mr. Bernard Traut, a technical assistant in those years, employed at the observatory from 1 December 1960.

Jean-Louis introduced me to Mr. Traut at my visit to Strasbourg in June 2013 and I talked with him for about an hour while he showed me e.g. the room with old instruments. One of the items I could immediately recognize as a model of a "complex mirror", the original name by Lacroute for his beam combiner, very similar to Fig. 2 in Kovalevsky (2009) where it is called a "multiple prism". Mr. Traut told me that Lacroute asked him to make such and such an item but never told him the purpose of the device. Not to him nor to his family did Lacroute ever mention that he worked on a satellite for astrometry, not before ESRO/ESA had become involved.

Quite naturally, I recently wrote to Jean-Louis who was my collaborator on the Tycho data reduction for many years (about 1985-1996).

**Jean-Louis Halbwachs** on 22 Oct. 2017

Dear Erik,

I received from Bernard Traut a short overview of his activities pertaining to the preparation of the Hipparcos programme :

1. In 1960-1961, Pierre Bacchus ordered the installation of an aluminium pipe 30 m long and 300 mm in diameter. This pipe was made of several sections, resting horizontally on an East-West axis. An objective and an ocular were mounted in order to observe in the meridian plane, and a few light sources were included. The pipe was used under vacuum in order to study the alteration of the images due to the atmosphere. This experiment was moved to Lille when Bacchus obtained a position of Professor of the university of Lille in 1963.

2. In 1968-1970, Pierre Lacroute has installed an optical device in front of the objective of the 21 cm refractor, in the southern dome of the observatory. The orientation of the device could rotate on 360°. According to Traut, the purpose was to search the most efficient basic angle for Hipparcos.

3. In 1970, Pierre Lacroute ordered several brass frames supporting spring blades. The blades were between 2 and 5 mm wide, and the interval between them was similar to the width. These frames were fixed on the optical device mentioned above, and also on an optical bench.

4. In 1970, Pierre Lacroute bought an oven ("étuve" in french) to make glass bonding tests. His purpose was to make the combiner mirror of Hipparcos.

5. In 1970, Pierre Lacroute bought 2 glass disks with a diameter of 400 mm and he asked Bernard Traut if he could go to Paris to work on them. Traut accepted, but the project has not been followed up.

The relation between Hipparcos and many of these points is not clear. Since Lacroute didn't explain why he needed these devices, Traut inferred a posteriori that it was for Hipparcos.

- Although it seems inconclusive, I know that Point 4 is obviously related to the preparation of Hipparcos: last Friday, Pascal Dubois confirmed that Lacroute presented to the staff of the observatory a seminar about the glue that could be used to make the combiner mirror of Hipparcos.

- As you certainly know, the most visible contribution of Bacchus and Lacroute to Hipparcos was their presentation to the IAU symposium 61 : 1974IAUS...61..277B. I asked to Dubois if he remembered something, but he answered that, aside from the seminar about the glue of the mirror, Lacroute didn't talk about his work. Therefore, we can infer from Point 4 and from his paper that he was certainly very concerned by the subject, but, unless spending a lot of time searching the archive, we cannot see the details of the genesis of the project.

Best regards,

  Jean-Louis

## A4  Letter from Daniel Egret

Hi Erik,

I am pleased you could find the thesis of Pierre Bacchus. I shall say I have nothing to add or to comment on these notes you have received from Jean-Louis Halbwachs and Bernard Traut and that I find very interesting to enlighten those years when space astrometry emerged. Obviously Pierre Lacroute did not share much about his projects with colleagues, at least at the beginning.

On my side, I joined the Observatory slightly after these episodes (in 1973) and came to the Hipparcos project much later, via the data processing issues, not through the astrometry itself.  At the Observatoire de Strasbourg, somebody else was involved in the astrometry aspect : Alain Fresneau. I know he was asked by Lacroute to work on some early designs of the mirrors, but he was also a student at  that time, and probably Lacroute did not tell him more.

I did not react to your other very comprehensive document of 20 October : feel free to use my inputs if you find them helpful. I agree with your statement that Lacroute and Kovalevsky were key persons without whom the project could not have emerged (and I agree also precisely on your statement that Lacroute can be considered as the father of space astrometry, not of Hipparcos).

I hope everything is well on your side. I met by chance Lennart in Paris Observatory gardens a couple of weeks ago, that was indeed a good surprise.

Cheers Daniel - on 30 October

## A5   Letter from Lennart Lindegren

Dear Erik,

Thank you for the draft. I read with great interest especially A3: these early activities of Lacroute are completely new to me.

I am not aware of any other written history of these early years.

Concerning Lacroute's ideas during 1975-1979, my knowledge is limited to the period from 1977 when all three of us were members of the ESA Space Astrometry Team, so I doubt that I can add anything that you did not already know. As you know, he was very keen on the possibility to apply dynamical smoothing in the data reductions and I remember having many discussions with him about this during meeting breaks. This was expected to improve the final astrometry by a significant factor (about 2 for faint stars, but much more for bright), but it required a smooth attitude, or at least only intermittent attitude control. I think this may have been for him the most important factor in deciding between different options.

I have been wondering if the idea of dynamical smoothing at least partly originated from his "synthesis" method for improving meridian circle observations, published in 1964 (Ann. Obs. Strasbourg, 6, 39), attached.

Regards,
Lennart    - on 30 October

**Notes by EH**:

1)  A search in ADS with "Lacroute 1964" gives five papers from Ann. Obs. Strasbourg, vol. 6, one of which is the one mentioned by Lennart.

2) Dynamical smoothing was first attempted in the Hipparcos data reduction, according to van Leeuwen (2007, chapter 10). It was only after the publication of the Hipparcos data in 1997 that these studies were further developed into a Fully Dynamic Attitude as it is called by van Leeuwen, resulting in much better precision and accuracy for the bright stars in the new Hipparcos catalogue of 2007.

## A6   References
**for the appendix**:

**2018.04.15**
*#2:The second of three reports on the early history of Hipparcos from 1964 to 1980*

# From TYCHO to Hipparcos 1975 to 1979

## Erik Høg

Niels Bohr Institute, Copenhagen University, Denmark     *ehoeg@hotmail.dk*

**Abstract:** With this report I follow the encouragement from several colleagues to my previous historical reports, they urged me to write more about these old times. I am going as deep as possible by means of documents, my own memory and in discussions by email with colleagues. - It is worth investigating how the Hipparcos astrometric satellite mission came about: was it almost a historical necessity that had to happen because astrophysicists needed the accurate positions, parallaxes and proper motion for the study of the Galaxy and the Universe and because the technical tools were available and affordable? Was it such general circumstances or did other more special historical circumstances play a major and even decisive role? My experience from 65 years dedicated to the development of astrometry shows me that special historical circumstances were needed and decisive.

## 1  Introduction

The historical circumstances in Western Europe in the 1960s and 1970s played a decisive role for the creation of Hipparcos, the first space astrometry mission ever, launched by ESA in 1989. This should become clear from the following where I describe and document some of the developments of Hipparcos from its beginning with the mission definition study in October 1975 up to the mission approval in 1980. A main focus will be on the development of the Options A and B, proposed by the present author and Pierre Lacroute, respectively, in December 1975. Their similarity and differences with respect to methods of measurement and scanning will be shown. Some of the many meetings are listed where the emerging space astrometry was presented and discussed with other colleagues. My archive of papers and technical notes is described.

I am placing my reports on arXiv and they are sometimes printed, hoping that they can be of interest for colleagues and historians. I do not intend to make a book although I have been urged to do so, a book of my scientific biography, but being 85 years old a book could put me under unwanted pressure to complete. Perhaps I could write an overview in a journal for the history of astronomy.

Below follow sections #2 - #6:
#2 an overview of the years 1975 to 1979,
#3 the Mission Definition Study - October 1975 to May 1976,
#4 meetings and Phase A Study - June 1976 to 1979,
#5 a brief conclusion, and
#6 references.

The Hipparcos space astrometry mission broke through to milliarcsecond (mas) astrometry for 100 000 stars, something impossible to do from the ground, and even in an absolute celestial coordinate system thus reaching or exceeding mas accuracy, not only precision. Ground-based astrometry could have thrived without Hipparcos by means of automatic meridian circles for absolute astrometry with perhaps 50 mas accuracy. Relative astrometry on the 50 mas level was obtained  with wide-field astrographs on photographic plates (like the Hamburg Zone Astrograph), but the mas accuracy could only be obtained in small fields of long-focus length, astrometric telescopes (e.g. parallaxes of a small number of targets).

That Hipparcos was affordable is astounding. At his famous talk in Prague in 1967, Pierre Lacroute mentioned a cost of 10 million French Franc for an astrometric space mission when the question of cost was asked. This was one of the reasons some listeners including myself considered the idea to be unrealistic, another reason was the primitive design. At a meeting on 18 December 1975 the AWG (Astronomy Working Group) of ESA assumed 15 MAU for space astrometry (MAU=Million Accounting Units = about a million Euro). In the end Hipparcos cost ESA about 500 MAU according to ESA (1997), i.e. almost the same as the famous Sydney Opera House. Even though such costs are not easily comparable over long intervals of time, it is clear that the original estimates were very much too low. For further information, the costs are 740 M Euro for the 5 year Gaia astrometry mission launched by ESA in 2013, and thanks to modern detector technology and larger mirrors Gaia is a factor one million times more efficient in the utilization of light from the stars than Hipparcos was.

The great vision by Lacroute about space astrometry was received with interest outside France, but it was a French project for ten years from 1964, and nobody outside France worked on it. The support in France came from several places and especially Jean Kovalevsky was leading. He was able to bring the project into ESA when it became clear that France could not do it alone, see Kovalevsky (2009). The French Space Agency had decided not to pursue any purely French scientific mission, but only support European programs. **Pierre Lacroute and Jean Kovalevsky were key persons without whom we would not have had the Hipparcos mission,** nobody was there who could have replaced them, as explained in EH2011b and Høg (2017f).

A symposium on Space Astrometry coordinated and chaired by Jean Kovalevsky, was held in Frascati, Italy, on 22-23 October 1974 and gathered 41 participants from Europe and the United States, see more in Kovalevsky (2009) and see in Sect. 3.5 my comment to my letter to C.A. Murray of 30 June 1975. This convinced ESRO (European Space Research Organization soon to be named ESA) about the scientific interest and it was decided to gather a small number of scientists for a mission definition study. I was invited and decided to join in spite of my profound scepticism. My deep scepticism about European space astrometry is explained in the letter of 30 June 1975, see Sect. 3.5.

The potential advantage of astrometric observation without the disturbing atmosphere was obvious, but the proposed instrumentation had held me from being interested. At the same time I was deeply involved in other big projects, developments of an automatic meridian circle and of a new type meridian circle, cf. Sect. 7 of Høg (2014). The paper by Requieme (1980) gives an overview of the time in that field of classsical astrometry. At this time astrometry was still being pursued at many observatories in Europe although astrophysics had long become the main research subject of astronomers. The natural consequence was that astrometry was gradually stopped and young astrophysicists were replacing astrometrists when they retired. **Thanks to this circumstance of active astrometry in Europe there was a basis on which sufficiently strong interest and enthusiasm for space astrometry could build** as soon as it began to look realistic, cf. EH2011b. This happened in Western Europe, but it could not happen in USA because astrophysics had been dominating for much longer thanks to large telescopes and good observing climate. In the USSR astrometry was strong, but political and economic conditions prevented to pursue space astrometry.

I have recently asked the opinion about the creation of Hipparcos from a number of colleague astronomers and from a historian of science, John L. Heilbron. The resulting twelve interviews are collected in Høg (2017f) to which the interested reader is referred. It was interesting and quite surprising for me to see that so different opinions existed on matters which had been clear and evident to me for years - and still are.

After the success of Hipparcos and the great interest generated by astrometric data, to this day the **astrophysical** community seems to not fully appreciate the decisive contributions toward the advent of space astronomy by e.g. Lennart Lindegren and myself. My own contributions during 65 years to the astrometric foundation of astrophysics are summarized in the poster Høg (2018a). A list of some of my papers since 2006 are given in Høg (2018b).

# 2   An overview of the years 1975 to 1979

This section is an adapted copy of Sect. A2 in Høg (2017f), for the reader's convenience.

Pierre Lacroute presented his ideas of space astrometry in Prague in Lacroute (1967) where I heard him, but I had never spoken with him before we met in October 1975 in Paris for the first meeting of the ESA Mission Definition Group. This meeting changed me from being very sceptical about space astrometry to become enthusiastic as explained and documented in Sect. 4 of Høg (2011b).

For the next meeting in December 1975 I proposed a new design with one-dimensional scanning, active attitude stabilization, revolving scanning mode, input catalogue etc., in all seven new features are listed in Høg (2011b), but there were in fact ten new features as now listed below in Sect. 3.1.

In January 1976, Pierre Lacroute invited me by the letter in Figure 1 to come to Observatoire de Paris before a study group meeting for a discussion between just us two, and he tells in the letter how much he appreciates that younger persons have now entered the project. We spent several hours in fruitful and pleasant exchange of ideas. Lacroute had immediately in December 1975 agreed to aim for a scanning satellite with one telescope, not with two telescopes as in the TD Option from Frascati, see ESRO (1975). The Spacelab option, the preferred option in the conclusions from Frascati, was still mentioned as a possibility in a note (see Høg 2018) in March 1976. He adopted an image-dissector tube (IDT) to become the very efficient primary detector behind a modulating grid instead of several photomultipliers behind long slits as in his original TD Option. In notes from December 1975 to March 1976 he also introduced a star mapper with slits and photomultipliers placed before the main field, able to detect stars and measure their position as required to point the IDT spot even without using an input catalogue. But his acceptance of other features in my proposal came more gradually and the chairmen of the group were obviously keen to avoid any decisions which might be premature or perhaps against Lacroute's ideas.

## 2.1   The development 1976 to 1979

I remember members of the study group saying that we were diverging because Lacroute maintained two-dimensional scanning with inclined slits, passive stabilization and no input catalogue since he considered this to be simpler and more safe. I assured them that we would soon converge towards my design.

But it took much longer than expected, nearly three years before the use of a modulation grid for two-dimensional scanning, a beam combiner in three parts, and passive scanning were definitively dropped. This appears from the fact that these options are still mentioned in ESA (1978), the "report on Phase A study", and in Barbieri & Bernacca (1979), the proceedings from the colloquium in Padua. Likewise according to the study report of 26 April 1978, the use of an input catalogue had not yet been decided. The final study report ESA (1979) does mention the input catalogue in Sect. 1.7.

The use of an input catalogue had, however, been decided already in a meeting of the Science Team on 23-24 November 1977 according to Høg (1997). I could therefore in December 1977 distribute an inquiry which I had held ready for some time on projects for the satellite mission to astronomers mainly in ESA countries and I went on a round trip giving lectures at a number of institutes to generate more interest and support for the mission project. This resulted in 80 projects of scientific investigation defined by about 50 astronomers at 12 institutions in ESA countries as reported in Høg (1979). These projects were analyzed by Høg (1979) and Turon Lacarrieu (1979) for their consequences on e.g. Galactic astrophysics.

The complete acceptance of one-dimensional scanning came after January 1979, according to Høg (1997 p. xxx=30). The same page contains further notes on the slow acceptance of the principles proposed in December 1975.

Lacroute's idea of attitude stabilization by the gravity gradient would conceivably work in the low-earth orbit originally assumed, but at the meeting of the science team in November 1977 we were directed by ESA to design the mission for a geostationary orbit which had become possible with the new ESA launcher Ariane. That killed the gravity gradient as an option for stabilization and active attitude control became obligatory. That also decided in favor of an input catalogue. According to Høg (1997), the main reason to keep the option with passive stabilization was its smooth rotation, while the active control with reaction wheels produced astrometric jitter of unknown size, and this question was never precisely answered. But the problem with reaction wheels was radically solved when MATRA introduced cold gas control. That came much later, during Phase B.

> The evolution of the Hipparcos project is described in ESA (1989). An account of the early phases of the Hipparcos project up to inclusion of the Tycho experiment in 1981 was presented in Venice as Høg (1997). The above text covers the period up to 1979 with a somewhat different emphasis. The early phase in the creation of Hipparcos and the mission selection process in 1980 are broadly described in Chapter 5 of the book by Perryman (2010). In the present report Høg (2018) I am describing my archive from those years focusing on the instrument design.

## 3   Mission Definition Study - October 1975 to May 1976

The first meeting of the Mission Definition Group (MDG) took place in Paris, i.e. in ESA Headquarters in Neuilly, on 14 October 1975. As mentioned in Sect. 4 of Høg (2011b), hereafter called EH2011b, my "profound scepticism and lack of interest in space techniques" was here changed to the opposite. The words of the chairman Dr. V. Manno that we should forget the existing proposals by Lacroute and just think how we could best use space technology for our science, these word were the "Sesame open!" for me.

Back in Denmark, at first I looked again at a completely different method than Lacroute, mentioned in the letter of 30 June in Høg (1975d) at the use of high-precision gyros attached to a telescope that could then be pointed at any star for measurement of its position. I mentioned it to students in my lecture on astrometry, but I soon turned my attention towards a scanning satellite. I remembered the Image Dissector Tube (IDT) and our brilliant electronics engineer, Ralph Florentin Nielsen, found information for me on this device which would be able to measure many stars one-by-one in the telescope field. That was the way forward.

The IDT was a technologically mature device, invented in the 1930s for television. CCDs also existed but I do not recall that I considered this very new technology, recently however I found in my archive for these years a reprint of the review by Samuelsson (1975).

Here follow the sections:
3.1  TYCHO / Option A by Erik Høg
3.2  TD / Option B by Pierre Lacroute
3.3  Options A and B down-select process
3.4  My archive for October 1975 to May 1976
3.5  Some quotes from my archive on the mission definition
3.6  The final study report from mission definition

## 3.1  TYCHO / Option A by Erik Høg

My work on space astrometry during the six weeks bore fruit, at first the 9 pages dated 2 December 1975 in Høg (1975a), but I had told of the ideas and distributed notes at a second meeting in Neuilly on 18 / 19 November. These 9 pages contain my first design of an astrometry satellite, then called TYCHO. The design was sent to Pierre Lacroute with a letter of 4 December 1975, Høg (1975b). The TYCHO design differed from that presented for a scanning astrometry satellite TD by Lacroute at the Frascati symposium in 1974, ESRO (1975) by introduction of **ten** new features for a scanning satellite - **not only seven** new features as counted by me in previous papers and lectures.

These ten new design features were implemented in the final Hipparcos satellite and mission:

1) use only one small telescope instead of two large telescopes as in the TD option by Lacroute,
2) use a beam combiner of two parts as in Figure 3a, not three parts as in Figure 2b,
3) use active attitude control,
4) make the spin axis revolve around the Sun at a constant angle,
5) use an Image Dissector Tube (IDT) as main detector,
6) use one-dimensional measurement along scan,
7) use a modulating grid instead of widely spaced slits,
8) use a star mapper with one photomultiplier to detect reference stars,
9) use an input catalogue with 600,000 selected stars, see Table 1 in Høg (1975a).

#1 meant that a scanning astrometry satellite with a telescope of 16x16 cm aperture could be launched by a Scout launcher instead of two telescopes with 40x40 cm and 27x27 cm apertures which required a Delta launcher.

In January 1976 I derived from the theory by Høg (1961) of division corrections on a meridian circle that:

10) the beam combiner should have a basic angle between the two fields of view other than the 90 degrees assumed by Lacroute, especially resonances at 90 and 60 degrees must be avoided. In fact anything else between 50 and 110 degrees will do with little variation of accuracy.

The number of stars in the input catalogue was reduced to 100,000 in March 1976. (For the actual conduct of Hipparcos observations a special Hipparcos Input Catalogue of 118,209 stars were selected in a large preparatory work before launch from some 214,000 distinct candidates contained in some 214 observations programs.)

All these ideas formed a self-consistent instrument which by mid-1976 looked as in the Figures 3a and 3b. See also the discussion in Høg (1997). **It appears from the historical context that this self-consistent design could hardly have been proposed at that crucial time by anyone else, in accordance with my previous claim in EH2011b and Høg (2017f).**

Following an advice from the chairman, I changed the name TYCHO to Option A or Astrometric Satellite (AS) which three years later became Hipparcos. Lacroute's new TD option also equipped with an image dissector tube became Option B.

Early phases of the Hipparcos project have been further outlined in the appendix of the recent report Høg (2017f).

## 3.2  TD / Option B by Pierre Lacroute

In response to my proposal in December 1975, Lacroute proposed already in the same month a new version of his Option TD, TD stands for Thor-Delta launcher, from 1973 which was later called Option B. It has only one telescope instead of two and an IDT is introduced as a much more efficient detector than photomultipliers. This is documented in the archive papers in the bunch "Lacr01".

The following Figure 4 is found In a report from 12 March 1976 showing how stars are detected and measured in two coordinates by Option B, a distinct difference from Option A with one-dimensional measurement. The Option B has here a beam combiner for three directions of view shown here in Fig. 4. This is mentioned in the final MDG study report (ESA 1976) in Sect. 2.3.2 as an option with some advantages but also with some problems in the diffraction pattern. The study report finally assumes a telescope of about 25 cm diameter with two directions of view obtained by a beam combiner with two reflecting surfaces, but this does not give the required symmetry for measuring in two dimensions.

Therefore in later reports from the Phase A study a symmetric beam combiner appears with three surfaces for two directions of view. The study report version #1 ESA (1978) specifies in Sect. 2.7.2  "the complex mirror is a triple split symmetrical device". The final report ESA (1979) specifies in Sect. 3.2.1 "the complex mirror consists of two mirrors, tilted in opposite direction" and Figure 3.2 shows accordingly a mirror with only two parts.

For Lacroute, however, the Option B from March 1976 with three directions of view appears to have been important since he maintains the concept after one year when it is further discussed during the Phase A study in a report Lacroute (1977) dated 25 January 1977. It is explained on p. 1 that the most precise smoothness of attitude motion and thereby higher accuracy of measurements will be obtained by having three directions of view. The report also discusses the computations, the feasibility of the design and finally proposes to have two IDTs.

Lacroute says that an important objection against Option A is that a pointing accuracy of 1 arcsec is required in order to point the photo cathode of the IDT on a star of the input catalogue and this will be very difficult to achieve, he says. Therefore the capability of Option B to observe without an input catalogue is important.

Option B was maintained in the study in parallel with Option A. The Option B has adopted five of the above 10 features for Option A: 1, 5, 7, 8, and 10, but not the other five: it uses passive attitude control, it does not use revolving scanning, it does not use an input catalogue, it makes two-dimensional measurements, it needs a symmetric beam combiner with three parts.

UNIVERSITÉ LOUIS-PASTEUR

**OBSERVATOIRE**

11, RUE DE L'UNIVERSITÉ
TÉLÉPHONE: 35.43.00
POSTES 265 - 266

STRASBOURG, LE 6 janvier 19 76

P. LACROUTE
à Monsieur Eric HØG

Mon cher Collègue,

Je vous adresse ci-joint une nouvelle étude sur le guidage pro­posé pour un balayage T.D. Comme je le demande à la fin, il serait bien utile, pour établir les éléments permettant de choisir un type de balaya­ge, T.D. ou Tycho, que soit étudié assez en détail comment on pourrait réaliser pratiquement le balayage Tycho.

J'ai bien reçu la documentation dissector et l'envoi des exem­plaires de votre projet méridien. Je vous en remercie.

J'apprécie vivement de ne pas être le seul à présenter des pro­jets, car ainsi il est possible de mieux discuter ce qui est possible et ce qui ne l'est pas. En outre, étant âgé, il est probable que je ne pour­rai suivre activement l'exécution des projets et la participation d'as­tronomes plus jeunes est très nécessaire.

Si j'ai présenté en détail certains projets, c'est pour essayer toujours de vérifier personnellement s'ils sont viables et pour soumettre mes projets à la critique de ceux qui sont plus habitués que nous aux ex­périences spatiales.

Je n'ai rien reçu encore de Monsieur WILSON qui devait étudier l'optique. C'est important. J'espère recevoir quelque chose de lui bientôt.

Lorsque vous aurez réfléchi aux papiers que je vous ai adressés et examiné vous-même certains problèmes, il serait probablement utile que nous puissions nous rencontrer assez tôt avant le 29 janvier pour, si possible arrêter des positions communes sur certains points avant la sé­ance. Peut-être avec Monsieur Wilson s'il donne signe de vie. Il est pro­bable que cela serait une bonne avance dans le travail du groupe.

Bien cordialement.

P. LACROUTE.

*= tidsnok*

**Figure 1** Letter of 6 Jan. 1976, one of many kind letters from Lacroute. He mentions that we should meet in person before 29 January and so we did.

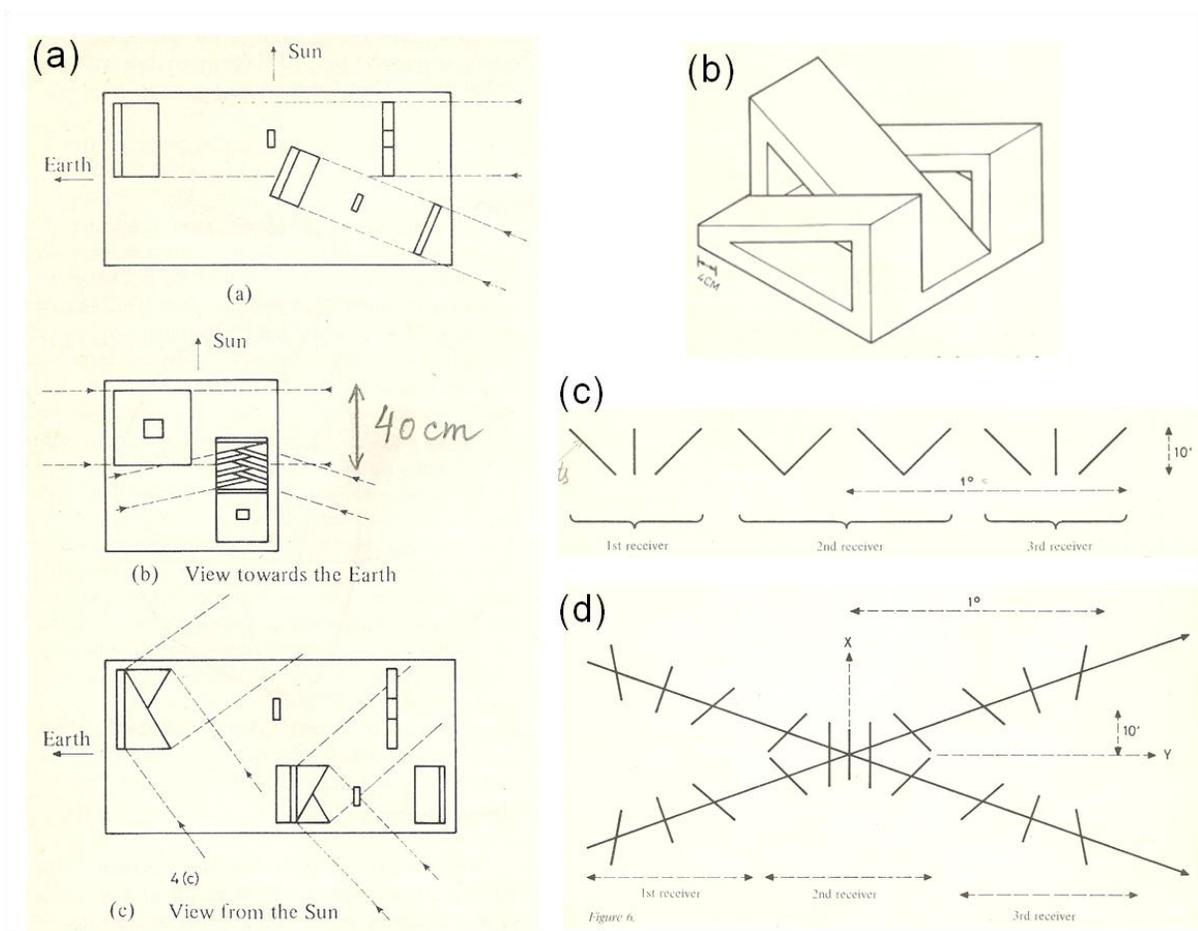

**Figure 2  (a)**: Two telescopes as proposed by Lacroute in 1974. By rotation about a spin axis pointing in the direction to the Sun the telescopes will continuously scan the sky with slits as in (c) and (d). **(b)**: A beam combiner placed in front of the telescope aperture will combine the beams from two fields on the sky separated by an angle of 90 degrees. The angle will be very stable as defined by the rigid material. **(c)** and **(d)**: The stars will cross the slits and be measured by six photomultipliers. The upper system is used in the larger telescope of (a), the lower one in the smaller telescope. Since the latter telescope is scanning a small circle on the sky the stars from the two fields move in different directions on the focal plane. Source: Lacroute (1974) and copied from EH2011b.

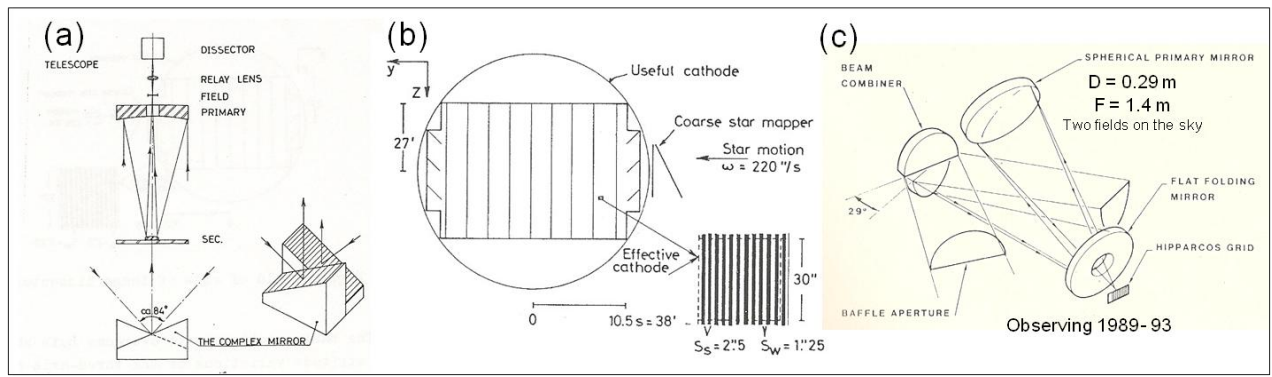

**Figure 3  (a)** and **(b)**: Hipparcos design according to Option A by mid 1976 (EH1977). **(c)**: When launched in 1989, the Hipparcos telescope was very different, a folded Schmidt system. The slit system and detectors were quite similar to (b), but the Tycho star mapper slits were implemented for the Tycho experiment proposed in 1981. Source: Copied from EH2011b.

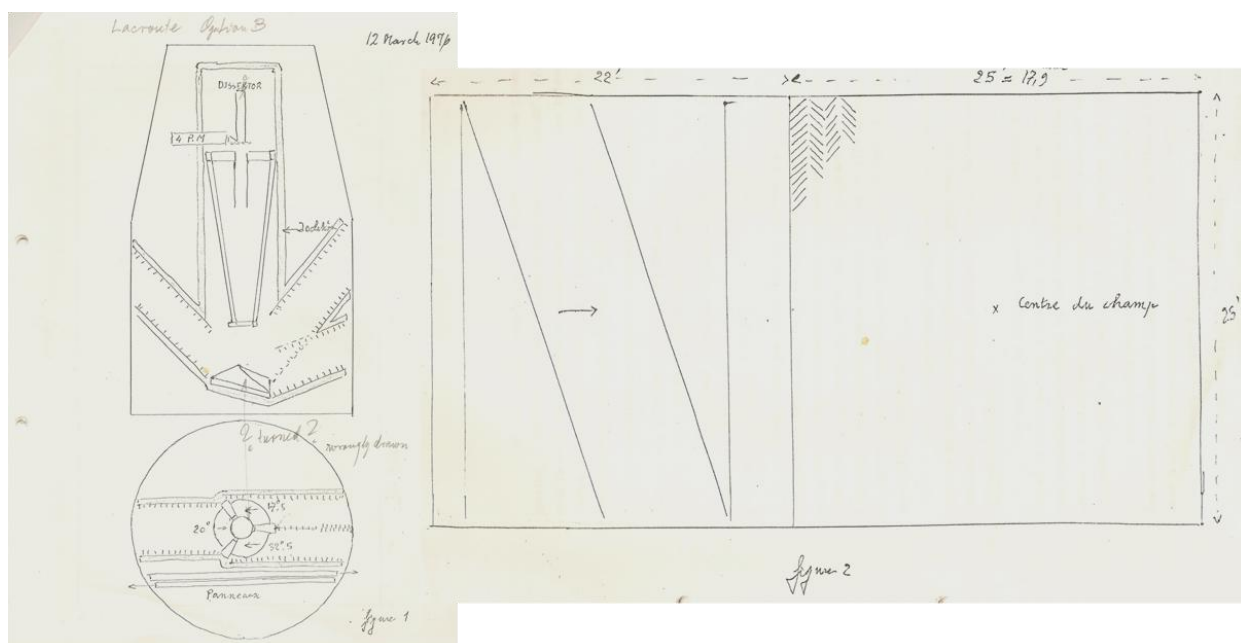

**Figure 4** Option B on 12 March 1976. **Left**: Beam combiner and baffles for three directions of view. **Right**: Focal grid common for the three fields of view. The star mapper with four slits allows observation without an input catalogue since the position of a star can be derived if it is detected with sufficient SNR at all four slit. Two-dimensional measurement can then follow in the IDT field at right. Source: Lacroute 1976.

## 3.3   Options A and B down-select process

These options were maintained in the study as having comparable astrometric performance and as subject to further studies which may result in the selection of one of the options or a combination of both, as the final study report from the mission definition says. In the end all ten features of Option A were implemented in Hipparcos.

A recent search in my archive has shown that the use of active attitude control and of an input catalogue were adopted at a meeting on 23-24 November 1977. Adoption of one-dimensional measurement and of a beam combiner with only two parts took longer, it had not yet happened in the first version of the Phase A study report ESA (1978) of 26 April 1978, but in the final report ESA (1979) of December 1979 it was there, just in time for the ESA committee meetings in 1980 where the fate of Hipparcos and other satellite projects would be decided. It would be an all-or-nothing decision - for Hipparcos it was ALL as told in Høg (2017d).

In the meantime many other issues were studied as shall be illustrated below, one of them was the immense task of data reduction which was developed assuming the Option A with revolving and one-dimensional scanning, especially after Lennart Lindegren joined us in September 1976.

## 3.4   My Archive for October 1975 to May 1976

My archive of reports and correspondence about the first period of ESA studies is placed in five boxes of 7 cm thickness. They are labeled by pencil: "Frascati 1974 Rumastrom. 1975-76", "ESA Historisk 1975-1976" and for the later period "ESA 77-81-82", "Phase A3", and "1979 AS. 8". My scientific archive will in due time be placed in the Royal Library in Copenhagen according to an agreement of 2010.

**Høg: Høg01**:

About 2 cm = 150 pp from Oct. 1975 to April 1976. E.g. the following:

The 9 pages dated 2.12.1975 in Høg (1975a) were "Input to MDG (Mission Definition Group) on Space Astronomy", received by ESA on 5 Dec. as the stamp on my copy shows. A copy was given to Catherine Turon in 2007 for the Paris archive.

A letter to the chairman of the MDG is contained in Høg (1975c). It is a 2-page report of my (first) visit to ESTEC on 10/11 Dec. mainly to discuss and learn about sensing, stability and control of the attitude of a spacecraft.

Many pages contain further work on Option A: The angle between axes i.e. the basic angle, optical systems, low-dispersion spectroscopy proposed, attitude control, attitude requirements cf. Sect. 3.4.5 in ESA (1976), data reduction, scientific objectives cf. Sect. 2.1 in ESA (1976), optical astrometry projects compared with FK4 = Table 3 in ESA (1976) on p. 18. Here are found my originals to the Figures 1, 2, 3, and 4 in ESA (1976).

**Lacroute: Lacr01:**

About 120 pages =15 mm received from Lacroute, stacked with the oldest at bottom. They are all typewritten including about 3 pages with drawings of focal plane and telescopes. Some are from 5 and 10 December 1975 in French. They are replaced on 1.01 1976, and 5.01 and 16.01, and Feb 1976. They are on the options Tycho, TD and SpaceLab. Then follow on 12 March translations to English ~20 pages.

**Jan. - April 1976:**

**R.N. Wilson:** 13 pp on optics

**W.N. Brouw**: 3 pp on rotational velocity analysis

**K.H. Davis**: 4+2 pp on baffling and on detectors

**P. Bacchus**: 7 pp on optics, in French

**M. Schuyer**: 10 pp on launch vehicle, system requirements, S/C configuration, S/C subsystems

From the other members of the MDG, **M.G. Fracastoro**, E. **Roth** and **Soo**, I have no papers in my files but I am sure they contributed to the study report.

**R. Pacault**: 3 pp on invitation to the fourth meeting on 8/9 April 1976 on drafting the final report of MDG, with list of members of MDG and table of contents and contributors: http://www.astro.ku.dk/~erik/xx/pacault2.pdf and p. 1 shown here as Figure 5 (pacault1.pdf)

**1975: H. Samuelsson**: 7 pp on Space application of CCD sensors. A review by Samuelsson (1975).

**1975**: **A. Boksenberg**: 2 pp on Television Detector Techniques. A reprint.

**11 June 1976 Copenhagen**, Colloquium on Space Astrometry. See Sect. 4.1.

**10 July 1976** Director of Planning and Futures Programmes of ESA: 3+6 pp. Open Letter to Space Scientists: Solicitation for Membership of Scientific Consultant Groups for: GRIST, SEOCS, Space Astrometry, and EXUV.

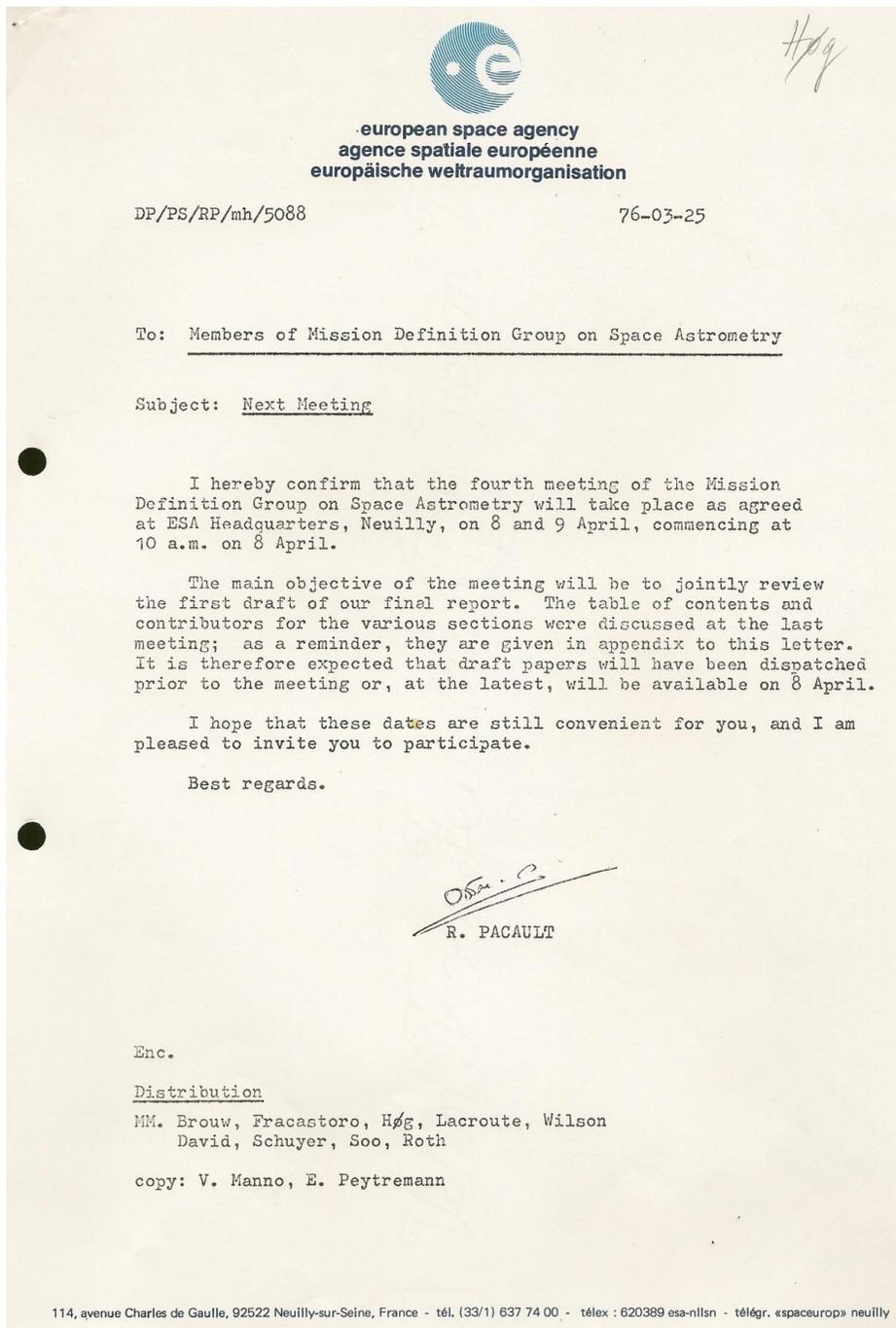

**Figure 5**  Invitation to the fourth meeting on 8/9 April 1976 on drafting the final report of the mission definition, distributed to the members of MDG as listed and including a table of contents and contributors.

The members of the Science Team were:

| | |
|---|---|
| C. Barbieri | Istituto di Astronomia, Padova (I) |
| F. Beeckmans | Visiting Scientist, Space Science Dept (ESA) |
| E. Høg | University Observatory, Copenhagen (DK) |
| J. Kovalevsky | CERGA, Grasse (F) |
| P. Lacroute | Dijon (F) |
| R.S. Le Poole | Sterrewacht, Leiden (NL) |
| L. Lindegren | Observatory, Lund (S) |
| C.A. Murray | Royal Greenwich Observatory, Herstmonceux (UK) |
| K. Poder | Geodetic Institute, Charlottenlund (DK) |
| F. Scandone | Florence (I) |
| S. Vaghi | Visiting Scientist, ESOC |

ESA staff members on the study were:

| | |
|---|---|
| H. Olthof | AWG Secretary |
| R. Pacault | Future Projects Study Office |
| E. Roth | Mathematical Analysis Division |
| M. Schuyer | System Engineering Department |

Requests for further information or additional copies of this report should be addressed to:

Dr. H. Olthof
Scientific Programme Directorate
European Space Agency
8-10, rue Mario-Nikis
F  75738 Paris Cedex 15.

**Figure 6**  Team members of the Phase A study according to the ESA study reports of 1978 and 1979.

## 3.5  Some quotes from my archive on the mission definition

25 October 1974: Høg (1974) 1p. Letter from Høg to R. Pacault giving credit for the arrangement of the Frascati Symposium and urging ESRO to study the feasibility of the proposed space astrometry projects. Copies of the letter were sent to Lacroute, Kovalevsky, Requieme, Strand, Fricke. - My concerns are stated in a letter of 30 June 1975, see below.

27 May 1975: ASTRO(75) 6, 26 pp on blue paper: Report on 14th Meeting of the Astrophysics Working Group held on 20-21 February 1975 at ESTEC, Noordwijk. p.15: Space Astrometry MS(74)36, 21 lines including a strong recommendation to perform a mission definition study as also recommended by the Solar System group. - In Frascati box.

30 June 1975: Høg (1975d) 2pp. Letter from Høg to Murray where I disagree with the conclusion in the report from Frascati. I explain my concern about "the technical feasibility of the proposed European Space Astrometry", the results as proposed by Lacroute in Frascati are not much better than can be obtained from the ground, "the effort on ground-based observations might decrease due to the great hopes attributed to space astrometry", "time is not ripe for space astrometry in Europe yet", and I mention my design of an Astrometric Space Telescope. My scepticism about space astrometry does not apply to the American astrometry project on the LST (Large Space Telescope). - But a few months later, with hindsight, I became grateful that the Frascati meeting had the **great leader Jean Kovalevsky** who was strong enough to write the conclusion and bring European space astrometry ideas into ESA.

1 October 1975: Letter from R. Pacault on the ESA Astrometry Mission Definition Study with a letter of appointment as a consultant to the group for the period 12 October 1975 to 31 Match 1976. Prior to this letter I received a phone call from ESA in Paris in which I was asked if I would join a study group. I agreed to join in spite of my scepticism because I considered it my duty to tell my opinion even if pessimistic or negative.

18 October 1975: My proposal for an Astrometric Space Telescope, 5 pp + 2 figures in Frascati box.

20.10 - 7.11.1975: My input to MDG Space Astrometry, 14 pages, incl. 6 figures, is placed in the Frascati box and was distributed at the second meeting on Friday **21 November**. Scientific arguments about reference stars; choice of payload where the scanning is compared with the spacelab version proposed by Lacroute and with a third alternative: a free-flyer with many maneuvres; complex mirror replaced with half-reflecting mirrors in two figures; on 3.11 a slit micrometer with image dissector is discussed and the revolving scanning is proposed, 4 figures are included, one of them showing a grid for two-dimensional scanning; on 7.11 UV/Visible Photometry with TYCHO is discussed.

12.11. 1975: FSP(75)16 from Neuilly, 4pp, Notes on the First meeting of MDG. Attendance: W. Brouw, E. Høg, P. Lacroute, R. Wilson, V. Manno and R. Pacault for ESA Headquarters, B. Morgenstern and M. Schuyer from ESTEC.

3 Dec. 1975 my proposal to ESA for a Shuttle-independent mission for astrometry with the TYCHO satellite described in Høg (1975a).

18 Dec. 1975 from AWG/Manno: mission profiles for the 1980-1990 period: SpaceLab experiment on Astrometry 15 MAU, if 500 MAU is assumed for all missions and nothing is allocated if 250 MAU is assumed. (MAU= Million Accounting Units).

17 Feb. 1976 my letter to Gibson, the Director General of ESA: My thanks for the invitation to become member of AWG. - My concern about long-term planning of both space astrometry and ground-based astrometry. I urge that not only the Galactic distance scale be mentioned in the ESA report (ASTRO (75) 16), stellar kinematics must be included, and I urge that space astrometry be called pioneering, not only exploitation. I was a member of AWG for three years and was succeeded by Andrew Murray.

23 January 1976: letter from R. Pacault on the third meeting of the study group, postponed from 29 February to take place on 19 and 20 February. The final report of the group is to be issued by the middle of April 1976.

5 April 1976: My arrival to Turin for lecture and discussion during three days at the observatory on invitation by prof. M.G. Fracastoro, member of the study group.

8 -9 April 1976: Fourth meeting of the study group in Neuilly.

## 3.6   The final study report from Mission Definition

It is available here as ESA (1976) and dated 25 January 1977. A preliminary version was issued in May 1976, available for the meetings in Copenhagen in June 1976. Some notes:

**Telescopes:** The telescopes for the two options are different in the final report because the required fields have different size, but the final Hipparcos telescope shown in Figure 3c was proposed much later by the industrial contractor MATRA during Phase B.

**Beam combiner:** For Option A the beam combiner is shown in Figure 1 consisting of two mirrors as in the final Hipparcos. I proposed in November 1975 an alternative beam combiner for Option A with full apertures for both fields obtained by a semi-transparent mirror in a pentagon arrangement, but this was dropped later on. R. N. Wilson comments in "third report on Optics" of 9 April 1976 on symmetry of the pupil. For Option B Figure 5 shows only two mirrors, thus without symmetry across scan as needed for the two-dimensional measurement. We will return to this problem in the Phase A study.

**Satellite and Orbit:** For the baseline options A or B a small spacecraft of 125 kg is launched by a Scout vehicle into a low sun-synchronous orbit at an altitude of 550 km according to the final report.

**Terminology evolved** in those years: Complex Mirror became Beam Combiner; Axes of Vision > Fields of View; Angle beteen Axes > Basic Angle; Guide Field > Star Mapper; Preselected Stars > Input Catalogue; TYCHO > Option A > Astrometric Satellite (AS) > Hipparcos.

**On the names TYCHO/Hipparcos**: I had in 1975 and also later proposed the name TYCHO for the mission but in vain. In 1981 however, a year after the Hipparcos mission had been approved in 1980, I proposed to take down all the photon counts from the star mapper slits in order to obtain astrometry from at least 400,000 stars. This was approved by the science team and by ESA and called the "Tycho experiment". It led us (Høg et al. 2000) to publish the Tycho-2 Catalogue with astrometry and two-colour photometry of the 2.5 million brightest stars of the sky. This meant that twenty times as many stars were catalogued as by the originally approved Hipparcos mission from 1980.

# 4  Phase A Study - June 1976 to 1979

Here follow in Sect. 4.1 a list of meetings before the beginning of Phase A, in Sect. 4.2 notes from the Phase A study itself, and in Sect 4.3 notes on special studies and International Meetings.

The Phase A study was carried out in the years 1977 and 1978 and the results were presented in two versions: Version 1 of 26 April 1978 in ESA (1978) and the final version of 26 December 1979 in ESA (1979).

## 4.1  Meetings June 1976 to Dec. 1976

10 June 1976 meeting of AWG in Copenhagen/Brorfelde: The proposal from AWG for astrometry is now 55 MAU. Astrometry is now placed under Conventional Satellite.

11 June 1976: Colloquium on Space Astrometry at Copenhagen University.
Rationale and program on 3 pp at ESA (1976a). List of 49 participants scanned to Colloquium (1976).

28-30 June 1976 in Paris: Symposium with presentation of Study Results for Future Scientific Missions
1 July Meeting of AWG to issue recommendations
2 July Meeting of SAC to make recommendations

30 August 1976 in Grenoble at the General Assembly (GA) of IAU: Joint meeting of Comm. 8 and 24 on astrometry. I presented: EH1977, Future astrometry from space and from the ground.

22 Sep. 1976 I had a meeting in Copenhagen with Lennart Lindegren, a Danish student, and a colleague where I explained the project and especially the challenging task of data analysis. This meeting is described in Høg (2008): **the great result of the meeting was that Lennart Lindegren became dedicated to space astrometry for the rest of his career.** Without Lennart Lindegren there would have been no Hipparcos mission approval in 1980, **in accordance with my previous claim in EH2011b and Høg (2017f).**

8/9 Dec. 1976 in Paris: Meeting of AWG. Teams for the feasibility studies (i.e. Phase A) were selected, the 12 members for astrometry were 7 astronomers and 5 from ESA, see the list of names in Figure 6 and a detailed list in: http://www.astro.ku.dk/~erik/xx/77AstromTeam.pdf.
More about the selection process on p. 3 in: Høg (2008).

## 4.2   Phase A study February 1977 to March 1978

According to ESA (1976) p.93, the Phase A study was recommended by the advisory committees in mid-1976 and decided by SPC in October 1976. A team set up in December, supported by ESA staff, was entrusted with the precise definition of the scientific specifications for the study. These studies were carried out from May 1977 to March 1978 by Airitalia and AML. During the latter part of 1978 and in 1979 the study was updated and refined by ESA staff and Matra.

**Input before the first meeting to be held on 15 Feb 1977** from members of the study team as requested by the chairman. This input description is based on ESA (1976)=DP/PS(76) 11, Rev.1 and papers in my archive.

19 Oct. 1976 from L. Lindegren, 7 pp: A three step procedure for deriving positions, proper motions, and parallaxes of stars observed by scanning great circles (Option A). Link given at LL2017.

2 Nov. 1976 from L. Lindegren, 8 pp: Relative mean errors of the five astrometric parameters. Link given at LL2017.

20 Nov. 1976 from L. Lindegren, 11 pp: The detailed equation of condition in Step 1. Link given at LL2017.

3 Dec. 1976 - 4 Jan. 1977 from E. Høg, 4+3+3+7+2+1=20 pp:
     Notes to Phase A of Space Astrometry
   1. Realization of an inertial frame
   2. Table 5. Mission models &
      Table 6. Comparison of scans
   3. Attitude control, Option A &
      Table 7 Attitude requirements
   4. Natural samples of nearby stars
      with Tables 8 and 9, Fig. 20
      Table 10. Sources of absolute proper motions
   5. Natural samples and selection effects
   6. Fig. 21. The roles of astrometric obs.

25 Jan. 1977 from P. Lacroute, 7 pp: Evolution of Option B, 3 field opt. system, improvement of the proposal from February 1976. Given at Lacroute (1977) with the file. See more above in the section on Option B.

8. Feb. 1977 from C.A. Murray, 10 pp:  He sees the greatest impact of a mission from the measurement of 6000 parallaxes for bright stars.

10 Feb. 1977 from C. Barbieri, 5+6 pp: Comments to Document ESA (1976)=DP/PS(76) 11, Rev.1. On p. 1: Opt. A seems to offer definite advantages because obs. limit as faint as 13 mag, but the seemingly much higher complexity of Opt. A over Opt. B gives worry. Repeating the mission after 10-15 years is mentioned here and in other reports as important for the accuracy of proper motions.

12 Feb. 1977 from L. Lindegren, 11 pp: The determination of the celestial sphere (Option A). Link given at LL2017. - A possible scheme for the reductions in Option A is sketched, the three-step method, with the purpose of getting a firmer basis for preliminary estimates of accuracies and for estimates of the required computing times.

14 Feb. 1977 from R. le Poole, 3 pp by hand: Comments to Document ESA (1976)=DP/PS(76) 11, Rev.1. On p. 1: Opt. A by far most attractive, particularly if revolving scanning can be implemented. Therefore detailed study of Opt. A. recommended keeping Opt. B as backup.

15 Feb. 1977 from J. Kovalevsky, 5 pp: Rotation of the satellite.

**Input after the first meeting held on 15 Feb 1977** from members of the study team

28 Feb. 1977 from L. Lindegren, 7 pp: On the possibility to measure parallactic displacements normal to the scan in Option A. Link given at LL2017. It is concluded that it will not be advantageous to introduce inclined slits. - This conclusion did not lead Lacroute to reconsider the inclined slits in Option B.

18 March 1977 from P. Lacroute, 2 pp by hand: Notes on the choice of grids. Grids for Option B with 3 fields discussed.

22 March 1977 from P. Lacroute, 4 pp by hand: Study on programmes for space astrometry.

21-22 March 1977 was the time of the meeting between C. Barbieri and R. le Poole, an undated report with 6+2 pp came shortly after: Comments on the proposed designs. Discussion of rectangular apertures versus half-mirror assembly assembly for the basic angle, concluding that the half-mirror is more attractive because of the symmetry. The Option B with three mirrors is omitted from discussion because it does not have rectangular apertures.

25 March 1977, from Simon J. Larcher, ACM, 30 pp: Proposal for theoretical study of the accuracy of an astrometric satellite.

31 March 1977 from ESA DP/PS(77)7, 17 pp: Technical specifications for Phase-A study of an astrometric satellite.

14+21 April 1977 from P. Lacroute, 5 pp: Comparison between option A and B.

27+28 April 1977: Second team meeting in Paris.

**23-24 November 1977: Meeting of the team.** Since the Ariane launcher had become available we were directed by ESA to adopt this as basis for our design, according to Høg (1997). The initial concept of a small satellite of 125 kg (yes, only 125 kg!) with a telescope of 20 cm aperture to be launched by a Scout vehicle into a low-earth orbit was therefore changed. Hipparcos became a larger spacecraft of 836 kg (see the mass analysis on p.58 in ESA 1979) with a telescope of 25 cm aperture to be launched into a near-geostationary orbit in a dual-launch with Ariane. This decision meant that only active attitude control could be used as proposed for Option A, no passive gravity gradient control as wanted for Option B would be feasible in this orbit far from the Earth.

At this same team meeting in November 1977 it was agreed to use an input catalogue, and the meeting agreed that I could launch an inquiry to astronomers about scientific projects. This inquiry by letters mostly to European astronomers is described in Høg (1979) and is mentioned in Sect. 1.7 of the final study report ESA (1979). The letters were followed up by visits to astronomical institutes in order to gain interest for space astrometry by discussion and colloquia. My travels in those years included Bochum, Munich, Heidelberg (ARI), Königstuhl (MPIA), Bonn, Hoher List, Herstmonceux (RGO), Hamburg, Aarhus and Washington.

There were other meetings of the science team but I have no details. !!!???

## 4.3   Studies and International Meetings 1978 and 1979

**Padua, Italy, June 5-7, 1978**: Colloquium on European satellite astrometry.

The proceedings by Barbieri & Bernacca (1979) list 47 participants and 32 contributions on 303 pages. NASA ADS gives a list of content with  authors and titles of the contributions but no abstracts.

**Study of Option A/Hipparcos:** In late 1978 a study was started in Copenhagen to determine whether accurate positions and parallaxes could be derived from one-dimensional observations with a scanning satellite like Hipparcos. The study made use of numerical simulations and the data reduction method proposed by Lindegren. The least-squares problem involving thousands of unknowns (astrometric data and attitude angles) was handled with a general geodetic adjustment program at the Danish Geodetic Institute. The good condition of the problem was confirmed and the results were cited in ESA (1979, p.75) and published by Høyer, Poder, Lindegren  & Høg (1981).

When asked about the time of this study Lindegren answered recently that it probably took place between December 1978 and October 1979 and he mentions two of his Technical Notes (TN) in Lindegren (2017). He wrote: The TN 1978-12-04 "Reconstruction of the celestial sphere. Suggestions for a numerical study of error propagation" could be the starting point, and the TN 1979-10-16 "Formulae for comparison of Copenhagen simulation results with theoretical predictions" could be near the end.

**Lennart Lindegren was at ESTEC** as a visiting scientist for eight months, starting in March 1979. He worked there on the final version of the Phase A study report ESA (1979) which is dated December 1979, just in time for the important meetings in January 1980 about the mission approval, reported in Høg (2017d).

**Montreal, Canada, August 14-23, 1979:** General Assembly of the IAU. There were presentations, but I have no details. In the Hipparcos Catalogue, I read: "The dialogue with the international scientific community was continued at special meetings: at the General Assemblies of the International Astronomical Union in Grenoble in 1976 and in Montreal in 1979, and at the Colloquium on 'European Satellite Astrometry' in Padua in 1978." No further details.

**Meetings at ESTEC, January 23/14 January, 1980:** The final decision by the AWG took place where Hipparcos was recommended and the EXUV project lost. This crucial meeting is described and discussed in Høg (2017d) where **Edward van den Heuvel is shown to be a person without whom there would have been no Hipparcos mission,** see especially my discussion with van den Heuvel on p. 8 in Høg (2017f). Ed as an X-ray astronomer was expected to support the EXUV mission, but he saw the much greater scientific value of Hipparcos and he was able to convince some other astrophysicists to vote for Hipparcos, enough to gain a majority in the AWG.

**Hipparcos Science Team:** Documents from the time after 1980 are listed in ESA (2018).

# 5  Conclusion

My notes from meetings are short, in pencil and hard to read. A much better source to the meetings exists in the extensive and nicely written notes in the hard cover protocols by Lennart Lindegren - if and when they become available.

I will stop my report here and leave the interested reader to consult the references and the official reports from the studies ESA (1976, 1978, 1979) and the overviews in ESA (1989) and the above Sect. 1.

**Acknowledgements:** I am grateful to Ulrich Bastian, Jos de Bruijne, Claus Fabricius, Povl Høyer, Lennart Lindegren, Jørgen Otzen Petersen, Knud Poder, Mattia Vaccari, Edward van den Heuvel, Andreas Wicenec, and Norbert Zacharias for encouragement to write this report and for comments to previous versions. From Jean Kovalevsky I received a very kind and deeply touching message on 19 February 2018: "Sorry, but I feel very tired. I did not read your paper and I trust you that everything is OK." I am also grateful for many comments from Jean during the past few months while I worked on the three reports on the early years of Hipparcos.

**2017.12.15**
*#3: The third of three reports on the early history of Hipparcos from 1964 to 1980*

### Miraculous 1980 for Hipparcos

*Erik Høg, Niels Bohr Institute, Copenhagen*

*ABSTRACT*: Many astrophysicists would agree that the astrometric foundation of astrophysics with positions, motions and distances of stars is important in all parts of astronomy and astrophysics. But in a situation where they have to chose between an astrometric and an astrophysical project the majority will chose the astrophysical, even after an outstanding astrophysicist has presented overwhelming arguments that the astrometric mission would be scientifically much more important. This extremely challenging situation became real at meetings in ESA on 23/24 January 1980 when Hipparcos stood against an EXUV project. This appears in detail from new documents which also show how a majority for Hipparcos was nevertheless gathered. Other documents from before 1980 on space astrometry are briefly described by Jean Kovalevsky, Lennart Lindegren and the present author (called EH hereafter) and links are given.

## 1. Introduction

The discussions in ESAs Astronomy Working Group (AWG) and the Science Advisory Committee (SAC) in 1979-80 have been summarised in the report Høg (2011) hereafter called EH2011. That report included information from documents and quotations from the witnesses Edward van den Heuvel, Jean Kovalevsky and Catherine Turon some of which is quoted in the following.

The evidence led to conclude in Section 5 that in case the approval would have failed, Hipparcos or a similar scanning astrometry mission would never have been realized, neither in Europe nor anywhere else. This conclusion still holds.

According to EH2011, I said to an astrophysicist in a coffee break in 1979 that Hipparcos must be approved now or it never will be because, "...*no matter how much you say you are impressed by space astrometry, in the end the majority would always put their own project higher*". This insight is confirmed by the present evidence where 5 astrophysicists voted for the astrophysical project, EXUV, in the face of glass-clear arguments in favour of Hipparcos presented by the astrophysicist Ed van den Heuvel.

Hipparcos prevailed thanks to a kind of miracle where the role of Ed van den Heuvel is of special interest. He was asked, as a past member of the AWG, to give presentations to the AWG and the SAC comparing both missions. He concluded that Hipparcos should be preferred over the competing EXUV mission for strong scientific reasons. In off-line talks before and during the meeting with some members of AWG he was able to convince two X-ray astronomers to vote for Hipparcos resulting in an 8 to 5 vote over EXUV in AWG. Other hurdles were encountered in the following meetings as reported in EH2011 where the complicated negotiations are laid out in detail.

Ed has recently found his handwritten personal notes to the two presentations and has allowed me to make them publicly available as van den Heuvel (1980 and 2017).

His recommendation of Hipparcos is most remarkable because an approval of Hipparcos meant a personal scientific sacrifice for him. His former boss Cornelis de Jager, chairman of AWG, had asked him to make the two presentations. After the AWG meeting, de Jager who had a big stake in the EXUV angrily said to Ed, *"what are you doing? You are killing your own project."*

In 1980, Ed at the age of 39 years was well into his career which led to outstanding academic positions distinguished through work on the formation and evolution of compact astrophysical objects such as neutron stars, black holes, and white dwarfs in binary systems, and for his investigation of gamma ray bursts.

## 2. Documents from before 1980

In EH2011, Lennart Lindegren is quoted for saying (in 2008) that he intends to write down the developments up to 1980 from his own perspective, but he cannot promiss a certain date. Asked in 2017 Lennart says, *"that date is still somewhere in the future (or more likely never). But it may interest you to know that I recently made (almost) all the Hipparcos Technical Notes from Lund available on the web."* - A link is here given as Lindegren (2017).

According to EH2011, Jean Kovalevsky will try to write before summer (of 2008) on the 1965-1975 period. This was made available as Kovalevsky (2009) through a link in Høg (2011b) and it is a reference in the present report. In Section 4 of EH2011, Jean mentions certain documents he has sent to me in 2008, but they cannot be found in my files at present.

The present author Erik Høg tells: As mentioned in Sec. 4 of Høg (2011b), hereafter called EH2011b, the meeting in Paris on 14 October changed my "profound scepticism and lack of interest in space techniques" to the opposite. The words of the chairman Dr. V. Manno that we should forget the existing proposals by Lacroute and just think how we could best use space technology for our science, these word were the "Sesame open!" for me.

My work on space astrometry during the subsequent six weeks bore fruit, at first the 9 pages in Høg (1975a). They contain my first design of an astrometry satellite, then called TYCHO. Following an advice from the chairman, this name was soon changed to Option A or Astrometric Satellite (AS) which three years later became Hipparcos. Lacroute's new TD option also equipped with image dissector became Option B. The new design of the satellite, Option A, by mid 1976 is shown in Fig. 3a,b of EH2011b and the seven new features are mentioned in this paper.

The 9 pages dated 2.12.1975 were "Input to MDG (Mission Definition Group) on Space Astronomy", received by ESA on 5 Dec. as the stamp shows. A copy was given to Catherine Turon in 2007 for the Paris archive.

A letter to the chairman of MDG is contained in Høg (1975b). It is a 2-page report of my (first) visit to ESTEC on 10/11 Dec. mainly to discuss and learn about sensing, stability and control of the attitude of a spacecraft.

Early phases of the Hipparcos project have been further described in the appendix of Høg (2017) and in Høg (2018)..

I will update the present report if further evidence of sufficient interest should become available.

## 3. Letter from Ed in August 2017

A handwritten letter is placed as van den Heuvel (2017). Though easily readable, it is quoted here *verbatim* for the reader's convenience.

"                                         Baarn, August 1, 2017

*Prof. Erik Høg*

*Dear Erik,*

*Herewith my personal notes at the SSAC+AWG meeting on 23 January 1980 and for my presentation on the subsequent AWG meeting, where I was asked to summarize the advantages and disadvantages of Hipparcos and XUV, I do not remember whether the AWG meeting was also on 23 January or on the 24-th of January 1980. Do you remember? [Final voting in AWG was on 24th.]*

*My notes consist of the following:*
*- Pages 1-8 are my notes of the presentation for SSAC+AWG (this includes a few pages of discussion at the end).*
*- Pages EXUV-1 to EXUV-5: the same for the XUV satellite,*
*- Pages A-E: Discussion in SSAC+AWG of both projects, compared to each other*
*- Pages H-1 to H-4: Notes which I made for myself afterwards, where I summarized what I remembered from the Discussions on pages A-E,*
*- Page H-5 which is my final conclusion about the advantages and disadvantages of Hipparcos and XUV, compared to each other;*

*I gave this summary in my presentation for the AWG, with the conclusion that Hipparcos wins.*

*In the AWG meeting I remember that I wrote on the white board basically what you see on page H-5.*

*Best wishes*
*   Ed van den Heuvel*
*da Costalaan 3*
*3743 HT Baarn*
*Netherlands*

*P.S. You may remember that in the final AWG vote it was the vote of X-ray astronomer Spada (from Bologna) that tipped the scale for Hipparcos. Spada, with whom I talked in the break before the voting session, then voted for Hipparcos."*

**Note by EH on 24 August 2017**: The votes became 8 to 5, so with one less Hipparcos would still have won with 7 to 6, but Schilizzi's vote for Hipparcos may also have been a result of Ed's convincing effort as described in EH2011 at the end of Section 3, *"The round of vote in AWG mentioned was in fact the final one on 24 January 1980 where the X-ray astronomer Spada voted for Hipparcos which would otherwise have lost to the EXUV mission. Also Dutch radio astronomer Schilizzi voted in favour. This gave the vote of 8 to 5 in favour of Hipparcos. Present at the meeting as members of AWG were thirteen persons: de Jager, Cesarsky, Delache, Drapatz,*

*Fabian, Grewing, Jamar, Murray, Perola, Puget, Schilizzi, Spada, and Swanenburg while Rego was unable to attend."*

*Among these thirteen persons, only Murray was not an astrophysicist; clear astrophysicists are: de Jager, Cesarsky (cosmic rays and the interstellar medium), Delache (solar physicist), Drapatz (optical instrumentation/interferometry), Fabian (X-astronomer), Grewing (millimetre radio astronomer/high energy astronomy), Jamar (UV space astrophysics), Perola (radio astronomer; High energy astrophysics), Puget (IR space astrophysics), Schilizzi (radio astronomy), Spada (X-ray astronomy, head of Space Research Lab, Bologna), Swanenburg (Gamma- and X-ray astronomy).*

## 4. The notes by Ed from 1980

The original notes by van den Heuvel (1980) on 25 pages are scanned and placed in four separate files corresponding to the parts described in the above letter.

They are named P1-8, PXUV, PA-E, and PH1-5 in the following where I list all of them and give literal reproduction of a few pages, but leave most without mentioning. My comments stand in square brackets [].

**P1-8:** Present. for AWG+SAC of Hipparcos proposal.
Ground based astrometry, global and local, are described leading to the question: What to be expected from Space?
HIPPARCHOS: Global, homogenous and absolute
ST (=HST) small-field relative astrometry
...

**PXUV:** Presentation for AWG+SAC of EXUV proposal.
....

**PA-E+PH0:** Discussion in SSAC+AWG.
...
**PHE**: ...
Make summary +
questions which we wish to ask:
**PH0**:
        For my AWG presentation [on 24 January]:
<u>SAC</u>    Criteria:
<u>1</u> Good science
but: high cost may rule out even good science.
<u>2</u> Continuity and exploitation of existing expertise
<u>3</u> <u>Mission</u>

Try to classify both projects acc. to the criteria.

**PH1-4:** Personal summary of discussions.
**PH1**:
Altogether! Summary for myself for AWG presentation

Summary of presentations and ensuing discussion

<u>Hipparchos - Scientific value is very great</u>
for problems ranging from:
- Homogenous fundam. ref. system for star positions of invaluable import. for future gener
- parallaxes:  Important for a variety of fundam. problems regarding stellar structure and evol.
-  Cosmic distance scale (cepheids, RRLyrae, abs. calib of man )
-  Prop. motions: dynamics of the galaxy (Fricke).
It was clear from the presentation and the subsequent remarks from the audience (one can hardly say: discussion), among which: Prof. Blaauw: puts a very much more solid basis under almost any of astron, Prof. Fricke, etc. - that there is a very great support from astron. community all over the world (12 IAU commissions).
Remark from Connes: ...??? [cf. Sect. 5]

The value for the astron. community will be very great (I think this is realized by everybody) and especially also for future generations. It is as fundamental as Bessel's work to obtain a first ref. system, and comparable to things like
    Bonner Durchmust., Cape Durchmust., Palomar Sky Survey - and has a value greatly exceeding only its simple astrometric importance.
    Questions posed from public:
...
**PH2**: [Quest. on Hipp. continued]
**PH3**: <u>EXUV:</u>
...
**PH4**: ...

**PH5:** Final conclusions.
For my AWG presentation:
Let us now consider these projects in terms of the criteria of Pinkau [see the above PB-E, the discussion in SSAC+AWG]:

[written on the white board:]

| A | Positive criteria | Hipparcos | EXUV |
|---|---|---|---|
| 1. | Good science | +++ | + |
| 2. | Continuity and exploitation of existing (Space) expertice in Europe | -- | + |
| 3. | Different in character from other agencies' programmes | +++ | -- |
| 4. | Missions for which European scientific Community expresses great interest and for which resources of industry and participating institutions offer adequate support | ++ | ++ |
| 5. | New and unique instrument or technique | ++ | **0** |
| | Summary | +++++++ | ++ |

| A | Negative criteria | Hipparcos | EXUV |
|---|---|---|---|
| 1. | Mission consumes large fraction of ESA's mandatory budget preventing ESA from reasonable diversity (criterion should be used carefully) | **--** | **--** |
| 2. | Mission too far ahead of time (not yet adequate technique) | **no** | **no** |
| 3. | Mission can be done more economically from ground (or rockets, balloon) | **no** | **no** |
| 4. | Missions for which technical and operational risks are too high | **no** | **no** |

<u>Conclusion</u>: Hipparcos <u>wins</u> on A
                 Both missions are equal on B

## 5. Recollections about 1980

The important presentation by van den Heuvel took place in an auditorium at ESTEC on 24 January 1980. Present were members of AWG, SSAC (then called SAC), the study team of Hipparcos to which I belonged and many others. I do not remember the presentation itself nor Ed van den Heuvel, but I remember that Walter Fricke spoke in the discussion, perhaps because he had not before shown much support for the Hipparcos project. At ARI, this project was not one of the official tasks in the 1970s, but Hans-Georg Walter should be mentioned as one of the supporters. Every time I saw Walter we spoke of Hipparcos and he never missed an opportunity to regret the lack of support it received from Fricke, probably because Fricke did not want to divert his institute from the construction of FK5. Why did Fricke then speak so strongly for Hipparcos at the meeting? Th. Schmidt Kaler told me many years later that he had been urging Fricke very much to do so.

Another memory from the meeting, stacks of the study reports for Hipparcos, EXUV and others were laid out to take. At one point I saw that there were very few in the Hipparcos stack. I asked an ESA man why they had made so few for Hipparcos, he answered that they had made the same number for all projects, but after the AWG meeting where Hipparcos had won the interest for Hipparcos had grown.

**Lennart Lindegren** has recently sent me the following from the meeting. Some background information: In 1977, Pierre Connes had proposed a very innovative and ingenious system for ground-based photoelectric measurement of (relative) parallaxes and detection of dark companions. In 1979 he published a Letter in A&A ("Should we go to space for parallaxes?", A&A 71, L1, 1979) where he argued that much progress was still possible in ground-based parallax work, and that the proposed astrometric satellite (Hipparcos) was not the best solution for all kinds of studies. This lead to an animated exchange of A&A Letters between Høg (A&A 75, L4, 1979) and Connes (A&A 76, L11, 1979), here a link is given to Connes (1979), where the pros and cons of ground-based observations versus space were discussed.

Concerning the ESA meeting in 1980, Lennart continued: In my recollection, after the presentation of Hipparcos, Pierre Connes stood up and said that you don't need to go to space to measure parallaxes. I do not remember his exact words, or to what extent he elaborated the point, but this was the gist of it. I had the impression that he was not at all against the satellite project, but thought that so much more could be done from the ground with improved techniques. At that meeting, however, his remark felt quite hostile.

Lennart ended with: My memory of all this is not very reliable, and Connes may have a different view of things.

**Jean Kovalevsky**: First, excuse me for not replying you before, but there were major difficulties with electronic links at the retirement house where I live now. I just recovered Internet.

I have sent you already several letters or texts. You may introduce any par you want. Only one remark. You rightly insist on the output of AWG in the process of deciding on the future mission, but it did not directly address the deciding power. There is an intermediary body which considers the proposals of both Working groups AWG and SSWG: it is the SAC (Science advisory committee) It is this body which addresses to and is heard by ESRO. And, as I wrote to you in my letter, the choice of Hipparcos was taken with the large majority of 5 votes against only one for the UV satellite. This step had more weight than the advice of the working group and your text should give more emphasis to this crucial 5/1 vote.

**Ed** then noted: Jean is fully right that the SAC was the body that had the power to make the final decision. But the SAC does this upon the advice of the AWG, which is an advisory body composed of scientists (specialists) in astrophysics and in Solar System physics.

The general policy is that the SAC follows the advice of the scientific priorities given by AWG in its specific domain. There is no clear way that this higher body could ignore the priority advice of these "lower" bodies of specialists. So, if the AWG had advised in favour of the EXUV mission, it is difficult to see how the SAC could not have decided in favour of that mission.

**Lennart** explains: SSAC means Space Science Advisory Committee. It is the committee above AWG and SSWG, and used to be called SAC in the Hipparcos days.

See p. 179 in ESA (2000), on p. 186 in the file containing the history of ESA.

**EH** notes: Seeing this "History of ESA" for the first time, I recommend to follow the tough process for the Hipparcos approval which is easy to do by a search in the file for Hipparcos with CTRL-F. On p. 168ff, Hipparcos, EXUV and Giotto/GEOS-3 are discussed.

**Catherine Cesarsky** sent this when I asked her: Unlike Ed, I did not keep notes, and it was a long time ago!

At the time, I was working on cosmic rays and the interstellar medium. Still, of these two missions, I remember well that I supported Hipparchos, I thought it was thoroughly new and original and a great première for Europe, but I don't remember if I intervened or what I said at the meeting, sorry.

At the same time, I feel a bit bad revealing this. When we are in those committees, we are not supposed to comment. But perhaps it is OK after so many years?

**Richard Schilizzi** on 28 Sept: I delayed replying until I'd had a chance to look through some of my archives to see if I could find any notes from my time in the AWG. So far nothing has turned up, but there is a possibility that some material is at ASTRON in Dwingeloo in the same boxes as all my QUASAT archives; I'll have a look when I next go there.

I do have a few memories about that AWG meeting. I was excited by the Hipparcos mission because it was innovative and the huge advance in astrometric accuracy would have been of fundamental importance. I'd spoken to Wim Brouw about the mission since he was in the ESA study team (as far as I remember), and there were the obvious overlaps with VLBI astrometry (AGN research with VLBI was my main research at the time). I'm sorry I don't remember talking to Ed at the AWG meeting (it was a long time ago) but there's no doubt we will have discussed the options.

One other memory. Cees de Jager had recently been appointed as AWG chair (following Giancarlo Setti, I think). However, he was not able to attend the presentations of both missions due to another commitment but that didn't stop him chairing the AWG meeting very efficiently and fairly. I remember being very impressed by that, knowing his preference for EXUV.

Jean and Ed have discussed the role of the SAC vs the Working Groups. In my view, it was important that the AWG vote was clear-cut in favour of Hipparcos (8-5 and not 7-6) as far as the SAC was concerned. If it had been 7-6, that could have been interpreted as essentially a tied vote which the SAC would have had to resolve.

**Andrew Fabian** on 29 Sept: I do not wish to contribute to a document which implies that those who did not vote for Hipparcos should be ashamed and were blind to glass-clear arguments. History writing should be left to those who can be impartial. I will merely correct Ed's implication that I was from Mullard (corrected by EH). This is only true in the sense that I did my PhD at MSSL, but I left in 1973 and had no direct involvement with space hardware thereafter.

**EH** on 16 Oct.: I have tried to contact others who were present in Jan. 1980 and for whom I could find an email address, Michael Grewing, a member of AWG, and also Pierre Connes, but none responded. I have also hoped that one of those voting for EXUV would tell the reasons, but in vain. - Later on: Grewing has answered, see Appendix B.

**EH** on 20 Oct.: The paper Connes (1979) gives a glimpse of a discussion mentioned above by Lennart Lindegren, quite interesting and a bit amusing in hindsight. The paper gives references to, e.g., three papers by Connes and two by me. See also the notes by van den Heuvel (1980, 5Heuvel1980Summ.pdf p1). As noted above, Connes addresses only the measurement of motions of stars, not of absolute positions which is the important capability of Hipparcos and Gaia.

## 6. Quotes about Hipparcos and EXUV

EH on 3 Oct.: Arguments around Hipparcos and EXUV have been quoted in Sect. 4 of EH2011 from the official reports of the AWG and SAC meetings. On astrometry for instance this in AWG: *"The Astrometry mission, HIPPARCOS, will give fundamental quantitative results to all branches of Astronomy. It emphasises typical European know how and will serve a community never before involved in space research"*; on the EXUV mission for instance this: *"The fact that the scientific objectives of this mission are being covered by two different missions proposed by other agencies (EUVE by NASA and ROBISAT by Germany) emphasises its timeliness."* This is an interesting fact but hardly a competitive argument for EXUV.

From the two SAC meetings in February: "Strong advocates for EXUV were also present at the SAC meeting: *"in the event that the Hipparcos payload would need to be funded within the mandatory programme, the SAC was divided as to whether Hipparcos should then remain the Agency's choice or EXUV should be carried out because this mission was considered by some members to be just as interesting."* (The quotation is literal, including spellings and emphasis.) In the end, Hipparcos was in fact financed within the mandatory programme."

Jean Kovalevsky's report in Sect. 4 from 2008 about the SAC meeting is also very interesting.

Finally as mentioned above in Sect. 1, my reasoning in Sect. 5 of EH2011: "In case the approval had failed" is what I think also today and begins with: "It appears that the approval could well have failed in which case I am sure Hipparcos would never have been realized. This proposition etc."

## 7. Ed van den Heuvel and Hipparcos

In Appendix A van den Heuvel gives the scientific reasons for becoming a strong supporter of the Hipparcos project. He ends with a very positive statement about the AWG which is however against the evidence given in this report.

The important role of **Ed van den Heuvel** is illuminated by the discussion in Høg (2017) beginning on 14 September.

**Acknowledgements:** I am grateful to Edward van den Heuvel for permitting his notes to be included here. I thank Ulrich Bastian, Catherine Cesarsky, Andrew Fabian, Claus Fabricius, Jean Kovalevsky, Lennart Lindegren, Richard Schilizzi, and Ed van den Heuvel for comments to earlier versions of this report.

# Appendix A

**Why I became a strong supporter of the Hipparcos
mission in the competition with the EXUV mission**

Edward P.J. van den Heuvel

My field of research basically is: stellar structure and
evolution. Since most stars are members of binary
systems, I specialized in the study of the evolution of
binaries. Through this I quite naturally moved into
studies of how neutron stars and black holes can form in
binary systems, and thus into: formation and evolution of
X-ray binaries and binary radio pulsars. So, I also
became a high-energy astrophysicist.

To test the correctness of stellar models, it is crucial to
compare the outcome of stellar structure calculations,
such as the luminosity of a star of a given mass and
chemical composition, with the observed luminosity of a
star of the same mass and composition. To obtain reliable
observed luminosities of stars, one must have accurately
determined distances of the stars. In 1980, no accurate
distances were available for O and B stars, since these
stars are rare and even the nearest ones are so distant that
accurate parallax determinations were not available. O
and B stars are just the ones that terminate life as a
neutron star or a black hole. Therefore, for understanding
their structure and evolution, and understanding the
formation of neutron stars and black holes, the Hipparcos
mission, which would be able to determine stellar
parallaxes with orders of magnitude higher accuracy,
would be of crucial importance. And this held not only
for the OB stars, as also for all other stellar types the
quantitative comparison of model predictions with the
observations would be tremendously improved by
Hipparcos.

For these reasons, I found Hipparcos a unique mission of
outstanding astrophysical importance.

It was unique, since nobody else in the world had come
up with the idea of an astrometric satellite. Here Europe
was in a unique position to make a huge step forward in
fundamental astrophysics.

This was not the case with the EXUV satellite, although
this also was a very nice project. Germany was already
working on ROBISAT (which later became ROSAT)
which had unique capabilities for observing  soft X-rays.
And in the US, NASA was planning an EUV (Extreme
Ultraviolet) mission. Together, these two missions could
do the same as the EXUV mission. So, the EXUV
mission was not as unique as Hipparcos.

However, a very great advantage that the EXUV mission
had over Hipparcos was, that in Europe there were
already a sizeable number of laboratories that had great
experience in building instrumentation for X-ray and UV
satellites: In Germany the Max Planck Institut für
Extraterrestische Physik, in the Netherlands the Utrecht
Space Research Laboratory, in Italy the Space Research
Laboratory of Bologna, in England, Mullard Space
Research Laboratory, etc. These laboratories could, with
partial funding from their own governments, supply
important instruments for the EXUV mission. No such
situation existed for Hipparcos, for which no laboratory
in Europe had any previous experience in instrument
development. For these reasons, contrary to the case of
EXUV, interest of the European space research laboratories in developing instrumentation for Hipparcos was basically non-existent.

On this basis, the expectation was that EXUV was a much easier mission to get approved in the AWG of ESA than Hipparcos. Also for ESA, EXUV was financially an easier mission, as the different participating countries would be able to carry part of the budget for this satellite. Also, several AWG members already came from the above-mentioned space research laboratories, e.g.: de Jager and Swanenburg from the Netherlands, Spada from Bologna, etc. On the other hand, there were only a few astrometrists in the AWG. Therefore, it was to be feared that Hipparcos would never be able to come through. The outcome of the vote about which mission should get the green light, would therefore depend strongly on the basic astrophysical arguments that could be put forward in favour of each of the two missions.

It was a wonderful sign of the scientific objectivity of the AWG, that in the end its members decided in favour of the best science, over the possible interests of their own laboratories, and chose for Hipparcos.

**EH on 31 Oct.:** Please also read the above Sect. 7.

# Appendix B

**Hipparcos 1980 in the AWG and SAC**

Michael Grewing - on 30 October

Dear Erik,

thank you very much indeed for sending the latest (as well as earlier) version(s) of your text, and the links to previous documents like your 2011 paper. In the meantime I have been able to read several of them.

This is a "last minute" reply because I was hoping to go back to my files from the 1980 period. My documents are, however, still distributed between 3 different places (1 in France, 2 in Germany), and I have at present only access to a small fraction of this material.

My general comment is, that the material which you have compiled makes very interesting reading, and it brings back many issues that mattered in the HIPPARCOS decision at the time. They are indeed worthwhile to memorize. The mission selection processes at ESA are still very complex, i.e. many considerations play a role beyond the scientific value of a mission, which should, of course, always be the Number 1 criterion. Your paper focusses on the role of the AWG and the (S)SAC. My question is, if it would not be worthwhile to also shed a bit of light on the broader context.

At mission selection, criteria such as e.g. the judgement of conceptual/technical matureness, the size of the community, the preparedness (and interest) of industry, the confidence in the cost estimate etc. all play a certain role. During the first rounds of selection, confirmation and approval, i.e. up to Phase C/D many of these criteria are still quite uncertain ("educated guesses") until they are progressively better understood during Phase A und especially Phase B.

Together with the science working group appointed for the mission, ESTEC plays a very important role during these different phases, and I think one of the reasons why HIPPARCOS has so successfully been implemented into the ESA program is owed to the excellent work done by the ESTEC group (and European industry) over many years!

Let me also mention one detailed aspect that was special to HIPPARCOS: during the 1970-ties and well into the 1980-ties, several attempts were made to increase the ESA science budget by raising the level of contributions to this mandatory program. This was very difficult to achieve because some of the nations did not want to compromise their own national space activities. A "way forward" was finally found by asking for national funding of payloads. This significantly increased the total level of funds that went into ESA's Science Program. Given the nature of the HIPPARCOS project, external payload funding would not have been a good idea, and it should not be forgotten that important national funding went into software development for the HIPPARCOS mission anyway.

I hope, the previous 2 paragraphs make it obvious that the voting by the AWG and the (S)SAC, while extremely important, represents only 2 steps in a much broader and longer decision making process which occured at SPC level to which the (S)SAC reports. At SPC level, HIPPARCOS' uniqueness certainly played strongly in its favor, as the competing proposal was faced with competion from scientifically similar proposals submitted to other agencies (e.g. in Germany and in the U.S.).

Before getting too long, I want to briefly clarify my own role during this time:

- I had been a member of the Science Working Group for the EXUV mission since 1977;

- I had been an ESA appointed member of the international IUE commissioning team since 1978, and did research in UV astronomy on interstellar and circumstellar medium questions (first with the COPERNICUS, then with the IUE satellite; I engaged in longer wavelengths projects (IR, mm) only later;

- I became a member of the AWG in 1979 (until 1983); the 1979/1980 selection was therefore my first such experience;

- I became German Delegate to the SPC in 1981 which followed the evolution of the HIPPARCOS project through all phases from its development and construction to its launch into space;

- at the end of 1981 I sent a telefax to Mike Perryman in which I offered support for the TYCHO project from the Astronomical Institute at Tübingen.

Should I recover my material from the 1980-ties period in the near future and find something that could be interesting to you, I will contact you again.

With my best wishes,

Michael

**EH on 31 Oct.:** I also recommend to follow the tough process for the Hipparcos approval in ESA (2000) which is easy to do by a search in the file for "Hipparcos" with CTRL-F. The discussion of Hipparcos, EXUV and Giotto/GEOS-3 is found on p. 168ff.